\documentstyle[11pt,epsf]{article}
\voffset = - 1.0in
\hoffset = - 1.0in
\def\lagr{\hbox{$\cal L$}}

\def\dalam{\hbox
{\vrule\vbox{\hrule\hbox to 1ex{ \hfill}\kern 1 ex\hrule}\vrule}}

\def\half{\hbox{$ {1 \over 2}$ }}
\def\i/h{{i \over \h}}
\def\1/2{\hbox{$ {1 \over 2}$ }}

\def\vf{\varphi}
\def\f{\phi} 
\def\p{\psi}
\def\bp{\bar \psi}

\def\E{\hbox{$\cal E $}}
\def\tE{\hbox{$\tilde {\cal E} $}}
\def\ve{\varepsilon}

\def\<{\langle}
\def\>{\rangle}

\def\ch{\cosh}
\def\sh{\sinh}

\def\h{\hbar}

\def\a{\alpha}
\def\b{\beta}
\def\g{\gamma}  
\def\d{\delta}  
\def\l{\lambda}   
\def\s{\sigma}
\def\r{\rho}  

\def\c{\chi}

\def\m{\mu}
\def\n{\nu}

\def\w{\omega}
\def\tt{\theta}

\def\({\left(}
\def\[{\left[}
\def\){\right)}
\def\]{\right]}

\def\inf{\infty}
\def\pd{\partial}

\def\sq{\sqrt{1+x^2}}
\def\tvf{\tilde \vf}

\title{\bf  Topological and non-topological solutions in
the 3-phase model of hybrid chiral bag} \author{ K.
Sveshnikov$^{a,b,}$\thanks{e-mail address:  costa@bog.msu.su } \ ,
\ Il.  Malakhov$^{a}$, \ M.  Khalili$^{a}$, \ S.  Fedorov$^{a}$
\\ {\em $^{a}$Physics Department and
 $^{b}$Institute of Theoretical MicroPhysics,} \\  {\em  Moscow State University,
Moscow 119899, Russia} }

\date{ \ \ \  }

\begin{document}

\maketitle


The 3-phase version of the hybrid chiral bag model, containing
the phase of asymptotic freedom, the hadronization phase as well as the
intermediate phase of constituent quarks, is proposed. For this model the
self-consistent solutions, which take into account the fermion
vacuum polarization effects, are found in (1+1) D. The renormalized
total energy of the bag is studied as a function of its geometry and
topological (baryon) number. It is shown that in the case of non-zero
topological charge there exists a set of configurations being the
local minima of the total energy of the bag and containing all the
three phases, while in the non-topological case the minimum of the
total energy of the bag corresponds to vanishing size of the
phase of asymptotic freedom.

\vskip 1.0in

\noindent
Keywords: Hybrid Chiral Bag Models, Solitons, Dirac Sea
Polarization Effects

\vfill

\eject

\section{Introduction}

At present the most perspective approach to the description of
the low-energy structure of baryons  is realized by the hybrid chiral
models (HCM) of quark bags [1-3]. The inital formulation of HCM
is given in terms of the little bag with massless quarks
and gluons confined in the phase of asymptotic freedom,
surrounded by the colorless purely mesonic phase, described by some
nonlinear theory like the Skyrme model
[3-5], while boundary conditions between the phases represent
the "chiral confinement" [6]. In (1+1) D these conditions coincide
with the exact bosonisation relations, what is the origin of
assumption, that the fermion theory
inside the bag and the meson one outside are actually
equivalent [7]. As a consequence, all the physical properties of the
bag should be independent of the choice of the boundary surface, what
is the essence of the Cheshire Cat Principle (CCP) [7,8].  However,
both the exact bosonisation and so the CCP can be rigorously proved
only in (1+1) D, whereas in (3+1) D even the  formulation
of such a problem  is ambiguous.  As a result, in the
(3+1) dimensional HCM based on the CCP there exists a rather small
set of observables (e.g. the topological charge), which really do not
depend on the bag radius [3,8].

 At the same time,  the
phenomenology of strong interactions predicts unambiguously the
existence of the characteristic confinement scale about 0.5 fm, and so
in the realistic (3+1) D models the CCP should be strongly violated
regardless on the proof of bosonisation. Moreover, in such 2-phase
HCM there is no place for massive constituent quarks, whose concept
has been shown to be very efficient in the hadronic spectroscopy [9].
From this point of view the most attractive situation is the one,
where the initially free, almost massless current quarks  transmute
firstly into ``dressed'' due to  interaction massive constituent
quarks carrying the same quantum numbers of color, flavor and spin,
and only afterwards there emerges the purely mesonic colorless phase.
The first step to such version of the bag is given by
the 3-phase chiral model, where the additional intermediate phase of
interacting quarks and mesons with non-zero radial size
is introduced [10,11]. This model allows to take self-consistently
into account: i) the phase of asymptotic
freedom with free massless quarks; ii) the phase of constituent
quarks, which acquire an effective mass due to the chirally invariant
interaction with the meson fields in the intermediate region of finite
size; iii) the hadronization phase, where the quark degrees of
freedom are completely suppressed, while the nonlinear dynamics of
meson fields leads to the appearance of the c-number boson condensate
in the form of a classical soliton solution, which provides the
topological nature of the model as well as the corresponding
quantum numbers.

It is worth-while to note, that the direct quark-meson
interaction is also considered in a number of other approaches to the
description of low-energy hadron structure, in particular,  in
the cloudy bag models [12-14], as well as in various versions of the
chiral quark-soliton models [15-19].  However, the role of
this interaction in each of these models is substantionally different.
In the cloudy bag models such  $\pi \bar q q $-coupling
is considered only perturbatively, while in
quark-soliton models it is the main nonlinear mechanism of
dynamical generation of the quark bag in the whole space.
In the case under consideration  an intermediate variant is
realized, where the contribution of the direct chiral quark-meson
coupling to the properties of the system is nonlinear, but
the confinement of quarks is provided by some additional
procedure. Such an approach allows to realize  the nonlinear
mechanism of dynamical mass generation in the intermediate region,
but unlike the quark-soliton models doesn't have  problems
related to the absence of the total confinement.

In the present paper a toy (1+1) D  model of this kind is considered,
in which in the intermediate region the one-flavor fermion field is
coupled in a chirally invariant way  to the real scalar field, which
possesses a nonlinear soliton solution in the exterior region. For
this model the self-consistent solutions  with different values of
topological charge are found by taking into account the effects of
fermion vacuum polarization. Within these solutions the renormalized
total energy of the bag is studied as the function of its geometry and
the topological charge of the solution.  It is shown that
for non-zero topological charge there exists a set of
configurations being the local minima of the total energy of
the bag and containing all the three phases, while in the
nontopological case the minimum of the
bag's energy corresponds to vanishing size of the phase of asymptotic
freedom.

\section{Lagrangian and equations of motion}

The division of space into separate phases is performed by means of the
system of subsidiary fields $\tt(x)$. To describe the underlying
machinery, let us consider the Lagrangian of the form $$\lagr_0 =
\half (\pd_{\mu}\f)^2 - \tt V(\f) + \half(\pd_{\mu}\tt)^2 - g_{0}^2
W(\tt), \eqno(2.1)$$ where the coupling constant $g_{0}$ of
self-interaction of the field $\tt$ is assumed to be large enough to
neglect the matter fields $\f$ in the dynamics of $\tt$ to the
leading order, and thereafter to use $\tt$ as background fields for
the dynamics of $\f$'s [10,20]. One can obviously insert in (2.1) as
many fields $\tt (x)$ as needed with the appropriate self-interaction
which will determine (almost) rectangular division of space into
regions, corresponding to different phases, while the
Lorentz-covariance will be broken only spontaneously, on the level of
solutions of equations of motion. Therefore in order to restore the
covariance one can use freely the framework of covariant group
variables [21].  Assuming further that subsidiary fields $\tt (x)$
have already formed the required bag configuration, let us start with
 the following Lagrangian:  $$\lagr = \bp i\hat\pd \p + \half
(\pd_{\mu}\vf)^2 -  \half m_{0}^2 \vf^2 \tt_{I} - {M \over 2} \[ \bp,
e^{ig \g_5 \vf} \p \]_{-} \tt_{II}  - \( {M_0 \over 2} \[\bp, e^{ig
\g_5 \vf}\p \]_{-}+V(\vf) \) \tt_{III} \ , \eqno(2.2)$$ with
$\tt_{I}=\tt (|x|<x_1) \ , \ \tt_{II}=\tt (x_1 \leq |x| \leq x_2) \ ,
\ \tt_{III}=\tt(|x|>x_2)$ being the step functions, selecting the
corresponding regions of the bag. The commutator over fermion fields
in terms, corresponding to the chiral fermion-boson coupling, ensures
the charge conjugation symmetry of the model. Note that in this model
the vacuum pressure term appears to be redundant, since due to the
existence of the intermediate phase the Dirac's sea polarization
behaves very specifically and itself yields  the required "inward
pressure".  Moreover, there is no special need in ``valence'' quarks,
since it is the boson condensate in the form of the topological
soliton which accounts now for the topological quantum numbers of
the bag.

So in initial stage we have the theory of two fields, the fermion
 $\p$ and the boson $\vf$. In the region  $I$ the
bosons have the mass $m_0$, in the region $II$ the field $\vf$ is
massless and interacts with fermions in a chirally invariant way,
what results in emergence of the effective fermion mass $M$.
In the region $III$ the effective fermion mass becomes equal to
$M_0$ and the self-interaction of $\vf$ switches on, providing the
appearance of a soliton solution for the boson field in this region.
To form the bag we  assume  the mass $M_0$ to be very large,
what leads to the dynamical suppression of the fermion field in the
exterior region. Proceeding further this way, we take the mass
$m_0$ infinitely large, then for convergence of the bag's energy
the boson field should vanish in the region $I$.  We'll
assume also that the self-interaction potential $V(\vf)$ is a
an even function. Then the solution of equations of
motion for the boson field could be either odd (the topological
charge is by convention equal to one), or even (the topological
charge is zero) function.

Let us consider now the behavior
of fields in more detail.  According to the general approach accepted
in hybrid models, we consider the boson field in the
mean-field approximation, i.e. it is assumed to be a $c$-number
field.  Neglecting temporarily the explicitly Lorentz-covariant
description, we will consider the center of mass system of the bag,
where $\vf(x)$ should be a stationary classical  field being
a background for evolution of fermions. In the region $I$ $\vf(x)=0$,
while in the region $III$ it decouples from fermions due to the
infinite effective mass of the latters and is formed uniquely by the
self-interaction $V(\vf)$.

The equations of motion read:

\noindent in the region $I$
$$ i\hat \pd \p=0 \ , \eqno(2.3a)$$
$$ \vf = 0 \ , \eqno(2.3b)$$

\noindent in the region $II$
$$ \(i\hat\pd - M e^{ig \g_5\vf } \)\p = 0 \ , \eqno(2.4a)$$
$$ \vf ''= ig {M \over 2} \< \[\bp, \g_5 e^
          {ig\g_5\vf}\p\]_{-}\> \ , \eqno(2.4b)$$

\noindent and in the region $III$
$$ \(i\hat \pd  - M_0e^{ig \g_5\vf }\)\p = 0 \ , \eqno(2.5a)$$
$$ -\vf''+ V'(\vf)=0 \ , \eqno(2.5b)$$
where $\< \ \ \ \>$ in Eq (2.4b) stands for the expectation value
with respect to the fermionic state of the bag. To simplify
calculations, we put further $g=1$, because the dependence on it can
be easily restored by means of the substitution $\vf \to \vf/g$. Then
the spectral problem for fermionic wavefunctions $\p_{\w}$ with
definite energy $\w$ reads $$ \w \p_{\w} = -i\a \p '_{\w}+ \b
e^{i\g_5\vf}\left[M \tt_{II} +M_0\tt_{III} \right]\p_{\w}.
\eqno(2.6)$$ Upon taking $M_0 \to \inf$, we get that $\p_\w \to
0$ in the region $III$ in such a way, that the term $M_0\p_\w$ in
eq.(2.6) vanishes, and the following boundary conditions at the
points $\pm x_2$ appear
$$\pm i\g^1 \p_{\w}(\pm x_2) + e^{i \g_5 \vf(\pm x_2) } \p_{\w}(\pm x_2) =
0 \ . \eqno(2.7)$$
 Supplied by the condition of continuity for $\p (x)$ on
boundaries between the regions $I$ and $II$, the eqs.(2.7)
provide  the correct formulation of the spectral problem for
fermions, confined in the bag.  Note, that the boundary conditions
(2.7) are actually the standard chiral boundary conditions for the
hybrid models [1-8].  However, they arise now as a direct
consequence of an infinite mass of fermions in the region $III$,
rather than from a local surface action, what is not completely
correct [10,22].  In the region $I$ the Eq.(2.6) is the equation
for free massless fermions $$ \w \p_I=-i\a \p'_I \ , \eqno(2.8)$$
while in the intermediate region $II$ one has $$ \w \p_{II}=-i \a
\p'_{II} + \b M e^{i\g_5 \vf} \p_{II} \ .  \eqno(2.9)$$ Conditions of
wavefunction's continuity on the boundary between $I$ and $II$ read
$$ \p_{I}(\pm x_1)= \p_{II}(\pm x_1) \ , \eqno(2.10)$$ while at
points $|x|=x_2$ the wave functions satisfy the boundary conditions
(2.7).  Meanwhile the field $\vf$ in the Eq.(2.9) is not arbitrary
but has to be determined self-consistently  from Eq.(2.4) with
corresponding continuity  conditions at points $|x|=x_{1,2}$.

\section{Solutions with non-zero topological
charge}

The essential feature of this bag configuration is the fact that the
coupled equations (2.4) in the closed intermediate region
$II$ of finite size $d=x_2-x_1$ possess simple and physically
meaningful solution which would be unacceptable if these equations
were considered in the infinite space. In order to obtain this
solution in the most consistent way, we perform firstly in the region
$II$ the chiral Skyrme rotation $$ \p_\w = \hbox{exp}(-i\g_5\vf /2)
\c_\w, \eqno(3.1) $$ whatafter the Eq. (2.9) and the boundary
conditions (2.7) transform correspondingly into $$ (\w-\half\vf')\c_{\w} = -i\a
\c'_{\w}+\b M\c_{\w}, \eqno(3.2)$$ $$ \pm i \g^1\c_{\w}(\pm x_2) +
\c_{\w} (\pm x_2)=0 \ . \eqno(3.3) $$ It follows from Eq.(3.2), that
if we assume the linear behavior for the field $\vf(x)$ in the region
$II$, namely $$ \vf'=const=2\l \ , \eqno(3.4)$$ then it
becomes the equation for free massive fermions $$ \n \c =-i \a \c' +
\b M \c \ , \eqno(3.5)$$ with eigenvalues $\n=\w-\l$ .  So the
fermions being massless in the region $I$, acquire the mass $M$ in
the region $II$ due to the coupling to the field $\vf$, whence the
intermediate phase emerges describing massive quasifree "constituent
quarks".

At this moment, the most
important feature of Eq.(3.5) is that it reveals
the sign symmetry $\n \to -\n$, which corresponds to the unitary
transformation of fermionic wavefunction $$ \c \to \tilde \c = i \g_1
\c \ , \eqno(3.6)$$ while  the chiral
currents $$ j_5=i \bp \g_5 e^{i\g_5 \f} \p =i \c^+ \g_1 \c
\eqno(3.7)$$ coincide for these sign-symmetric states:  $$ j_5=i \c^+
\g_1 \c = i {\tilde \c}^+ \g_1 \tilde \c = \tilde j_5 \ .
\eqno(3.8)$$
However, the sign symmetry of Eq.(3.5) itself cannot provide
the corresponding one for the fermion spectrum, since
it takes place in the region $II$ only, while the latter
has to be determined from the Dirac equation on the unification of
the regions $I+II$. Meanwhile in the region $I$ one has the Eq.(2.8),
which possesses another symmetry, namely $ \w \leftrightarrow -\w $.
That means that the sign symmetry $ \n \leftrightarrow -\n $ of
the fermionic spectrum could hold only for discrete values of the
derivative $\vf'$ in the region $II$. These values should be
determined from the transcendent algebraic equation for fermionic
energy levels, which is obtained from the straightforward solution of
Eqs. (2.8) and (2.9) with account of boundary conditions (2.7) and
the constraint (2.10), and reads $$ \exp(4i \w x_1 )= { 1- e^{-2ikd}
{M-i (\n+k) \over M- i(\n-k) } \over  1- e^{-2ikd} {M+i (\n-k) \over
M+ i(\n+k) } }  \ { 1- e^{2ikd}  {M-i (\n-k) \over M- i(\n+k) } \over
1- e^{2ikd} {M+i (\n+k) \over M+ i(\n-k) } } \ , \eqno(3.9)$$ where
$\n^2=k^2+M^2 \ .$ Analysing the Eq. (3.9), one easily finds, that
the fermionic spectrum reveals the symmetry $ \n \leftrightarrow -\n
$, if $$ 4\l x_1 =\pi s \ , \eqno(3.10)$$ where $s$ is integer, since
for such values of $\vf'(x)$ in the region $II$ the l.h.s. of Eq.
 (3.9) reduces to $(-1)^s \ \exp (4i\n x_1) $.

Assuming the condition (3.10) to hold, the following consequence of
arguments becomes reasonable. In the r.h.s. of Eq.(2.4b), which
determines $\vf''(x)$ in the region $II$, we have the v.e.v.  of the
$C$-odd chiral current $$ J_5=\1/2 \[ \bp , i\g_5 e^{i\g_5 \f} \p
\]_-=\1/2 \[ \c^+ , i\g_1 \c \]_- \ , \eqno(3.11)$$ with $\c$ being
now  the secondary quantized Dirac field in the chiral representation
(3.1) $$ \c (x,t)= \sum \limits_n  b_n \c_n(x) e^{-i\w_n t} \ ,
\eqno(3.12)$$ where $\c_n(x)$ are the normalized solutions of the
corresponding Dirac equation, $b_n \ , b^+_n$ are fermionic
creation-annihilation operators which obey the canonical
anticommutation relations $$ \{ b_n , b^+_{n'} \}_+= \d_{nn'} \ ,
\quad  \{ b_n , b_{n'} \}_+=0 \ .  \eqno(3.13) $$ The average over
the given bag's state includes, by definition, the average over the
filled sea of negative energy states $\w_n <0 $ plus possible filled
valence fermion states with $\w_n >0 $ which are dropped for the
moment because their status is discussed specially below. Finally, $$
\< J_5 \>=\< J_5 \>_{sea}= \1/2 \( \sum \limits_{\w_n <0} -  \sum
\limits_{\w_n >0} \) \ \c^+_n i\g_1 \c_n  \ .  \eqno(3.14)$$ Let us
emphasize here, that in Eq.  (3.14) the division of fermions into sea
and valence ones is made in correspondence with the sign of their
eigen-frequencies $\w_n$, which differ from sign-symmetric $\n_n$ by
the shift in $\l$ $$ \w_n=\n_n+ \l \ , \eqno(3.15)$$ and so do not
possess the sign symmetry $ \w \leftrightarrow -\w $. However, if we
suppose additionally, that $\n_n$ and $\l$ are such that for all $n$
the signs of $\n_n$ and $\w_n$ coincide, i.e. after shifting by $\l$
none of $\n_n$'s changes its sign, then the condition $ \w_n {> \atop
<}  0$ in Eq. (3.14) will be equivalent to the condition $ \n_n {>
\atop <}  0$ .  Hence $$ \< J_5 \>_{sea}= \1/2 \( \sum \limits_{\n_n
<0} - \sum \limits_{\n_n >0} \) \ \c^+_n i\g_1 \c_n =0 \eqno(3.16)$$
by virtue of relation (3.8). In turn, it means that (2.4b) in the
 region $II$ reduces to $\vf''=0$, what is in complete agreement with
our initial assumption that $\vf'(x)=const$ in the region $II$. In
other words, we obtain the solution of the coupled Eqs. (2.4) in the
region $II$ in the form $$ \vf(x)= \Bigg\{ {2 \l (x-x_1) \ , \quad \
\ x_1 \leq x \leq x_2 \ , \atop \ \ 2\l (x+x_1) \ , \quad -x_2 \leq x
\leq -x_1 \ , } \eqno(3.17)$$ where $\l$ takes discrete
values from (3.10) and the fermion energy spectrum is determined from
the relation (3.15), while $\n_n$ is defined from Eq.  (3.9) after
replacing  the l.h.s. to $(-1)^s \ \exp (4i\n x_1) $.

There are the following keypoints that make this solution meaningful. The
first  is the finiteness of the intermediate region size $d$, because
for an infinite region $II$ the solution (3.17) should be
unacceptable. In our case, however, the size
of the intermediate region is always finite by construction, while the
boson field $\vf(x)$ acquires the solitonic behavior in the region
$III$ due to self-interaction $V(\vf)$.  Here the following
circumstance manifests again: in (1+1)D the chiral coupling $\bp
e^{i\g_5 \vf} \p$ itself cannot cause the solitonic behavior of the
scalar field by virtue of the effects of fermion-vacuum polarization only,
i.e. without additional self-interaction of bosons [23].
The second point is the discreteness and  the $ \n \leftrightarrow
-\n $ symmetry of the fermionic spectrum, what  leads in  turn to
a reasonable method of calculation for the average of the chiral
current $J_5$ over filled Dirac's sea (3.16), as well as for other
$C$-odd observables like the total fermion number.  After all, in the
case we consider the boson field is continuous everywhere and so is
topologically equivalent to that odd soliton, which would take place
in absence of fermions due to the self-interaction $V(\vf)$ only.
That's why the topological number of the boson field doesn't depend
on the existence and sizes of the spatial regions containing fermions
(the regions $I$ and $II$). On the other hand, the baryon number of
the hybrid bag is, by definition,  the sum of the topological charge
of the boson soliton and the fermion number of the bag interior. In
our case the latter is  zero, hence the baryon number of the bag is
determined by the topological charge of the boson field only and
doesn't depend on the sizes of the regions $I$ and $II$ containing
fermions, what is in agreement with the general ideology of hybrid
models.  More detailed discussion of this solution of Eq.(2.4) and
the arguments in favor of its uniqueness are given in Ref.[10].

It is also worth-while noticing, that although the
(topological) quantum numbers of such a bag  are determined
by its solitonic component, it doesn't mean that the filled
fermion levels with positive energy shouldn't exist at all. This
could take place for small enough values of the
parameter $\l$ only. If $\l$ increases, the negative levels $\w_n= -
|\n_n| + \l $ will inevitably move into the positive part of the
spectrum. The change of sign of each such level will decrease
$\<Q\>_{sea}$ by one unit of charge, but if we fill the emerging
positive level with the valence fermion, then the sum
$Q_{val}+Q_{sea}$ remains unchanged.  Analogously, the total axial
current will be equal to $J_{val}+J_{sea}$ and won't change
either, what ensures the vanishing r.h.s. of Eq.
(2.4b) and so preserves the status of linear function (3.17) as the
self-consistent solution of the field equations. Therefore, the
existence or absence of valence fermions in such
construction of the ground state of the bag depends actually on the
relation between $\l$  and $|\n|_{min}$ and so appears to be a
dynamical quantity like the other parameters (the size and mass)
which are determined from the total energy minimization procedure.

Another essential feature of this bag configuration is that (3.17)
provides the self-consistent solution of Eqs.(2.4) for even values
$s=2r$ in (3.10) only. The reason  is that for odd
values  $s=2r+1$ the fermionic spectrum obtained from the solution
of Eqs.(2.7-10) under condition $\vf'=2\l$ will always contain the
nondegenerate energy level $\c_0(x)$ with zero frequency
$\n_0=0$, while for even values  $s=2r$ all the $\n_n \not =0$.
According to the general theory [24],  such a
zero mode yields fractionalization, what means,
that its contribution   to all C-odd
observables is given by the operator $\1/2 (b_0^+b_0-b_0 b_0^+)$
with the eigenvalues $\pm 1/2$ and the numeric coefficient determined
by $\c_0(x)$. Now let us note,  that in
Eq.(2.4b) the chiral current should be averaged over its eigenvector
in order to provide the vanishing dispersion of the r.h.s., otherwise
the system of Eqs.(2.4) would be ill-defined.
So the operator part of the zero mode contribution to the
r.h.s. of (2.4b) turns unavoidably into the factor $\pm 1/2$,
meanwhile the wavefunction $\c_0(x)$ appears to be such that the
corresponding chiral current  in the region $II$ does not vanish (it
is proportional to $ \exp(-2M|x|)$). Hence the r.h.s. of Eq. (2.4b)
doesn't vanish for odd values  $s=2r+1$, and the function (3.17) is
no longer the self-consistent solution of Eqs. (2.4).

It's
not difficult, however, to find the way of constructing analogous
bags, where the odd values $s=2r+1$ are allowed. This method is based
on a specific for such two-dimensional bag models possibility to
choose in the model Lagrangian independently the signs of the chiral fermionic
masses $M,M_0$, both to the right and to the left of
the region of asymptotic freedom. More specifically, the considered
configurations with even $s$ appear from the Lagrangian (2.2),
which is symmetric in signs of fermion masses
 to the left and to the right.  Now let us consider the
situation, when in the Lagrangian (2.2) all the chiral masses to the
left of the central region change their signs. As a
result, we obtain the following Lagrangian
 $$\lagr = \bp i\hat\pd \p + \half
(\pd_{\mu}\vf)^2 - \half m_{0}^2 \vf^2 \tt_{I} - {M \over 2} \[ \bp,
e^{ig \g_5 \vf} \p \]_{-} \( \tt_{II}^{(+)}  - \tt_{II}^{(-)} \) - $$
$$ - {M_0 \over 2} \[\bp, e^{ig \g_5 \vf}\p \]_{-}  \(
\tt_{III}^{(+)} - \tt_{III}^{(-)} \) - V(\vf) \(
\tt_{III}^{(+)} + \tt_{III}^{(-)} \) \ ,
\eqno(3.18)$$ where $\tt_{I}=\tt (|x|<x_1) \ , \
\tt_{II}^{(\pm)}=\tt (x_1 \leq \pm x \leq x_2) \ , \
\tt_{III}^{(\pm)}=\tt( \pm x > x_2) \ , $
 which is antisymmetric in signs of the fermionic masses to the
left/right. Qualitively their behaviour for these two cases is
shown on the Fig. 1.

\vskip 3 true mm
\epsfbox{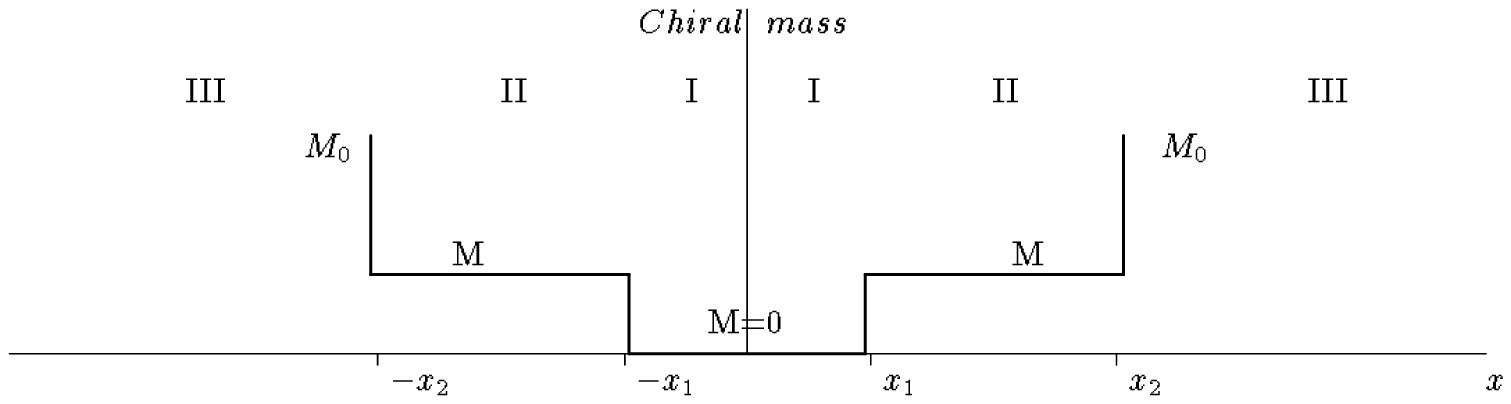}

\epsfbox{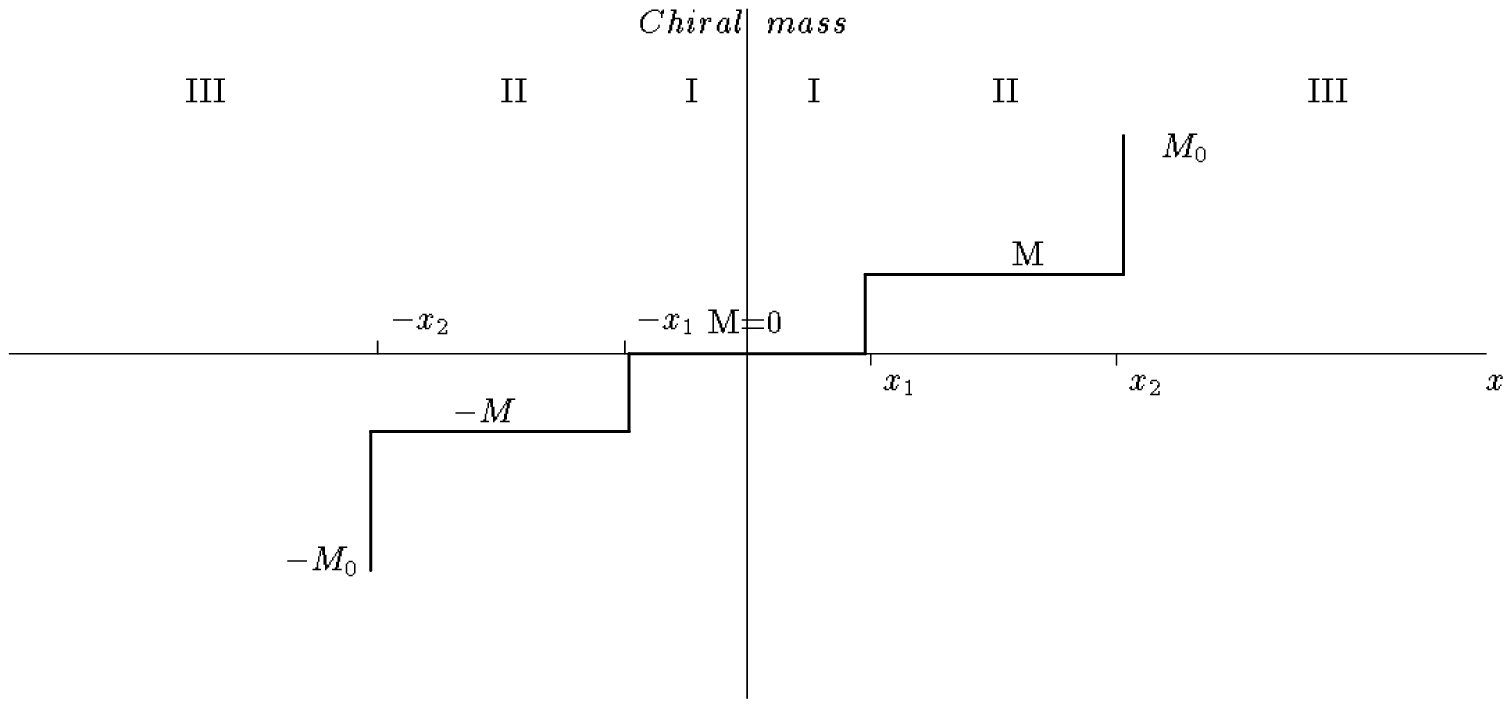}

\noindent Fig.1. The behaviour of the chiral fermionic masses for the
symmetric (up) and antisymmetric (down) cases.

\vskip 5 true mm

\noindent
It's easy to see, that the change of the sign of $M_0$ to the
left leads  merely to the change of the sign in the corresponding
boundary condition. Namely, in this case  we will have at the
point $x=-x_2$ $$ i\g^1 \p_{\w}(- x_2) + e^{i \g_5 \vf(- x_2) }
\p_{\w}(- x_2) = 0 \ , \eqno(3.19)$$ instead of (2.7). In the
intermediate region after the Skyrme rotation the Dirac equation
(3.5) turns into $$ \n \c =-i \a \c' \pm \b M \c \ , \eqno(3.20)$$
where in the r.h.s. of (3.20) the sign $\pm$ corresponds to the sign
of the chiral interaction to the right/left. In this
case again the $\n \to - \n$ symmetry of the spectrum  takes
place for the transformations $ \c \to \tilde \c = \s_2 \c $,
which do not alter the chiral current $ j_5= \tilde j_5 $, and so
the solution based on the linear Ansatz (3.4) and $\w = \n + \l $
remains valid, while the spectral equation for fermionic
eigenfrequencies appears firstly in the form $$ \exp(4i \w x_1 )= - \
{ 1- e^{-2ikd} {M-i (\n+k) \over M- i(\n-k) } \over  1- e^{-2ikd}
{M+i (\n-k) \over M+ i(\n+k) } }  \ { 1- e^{2ikd} {M-i (\n-k) \over
M- i(\n+k) } \over 1- e^{2ikd} {M+i (\n+k) \over M+ i(\n-k) } } \ ,
\eqno(3.21)$$ and differs from the Eq. (3.9) by the sign in the
r.h.s.  The origin of this additional sign is
the antisymmetry of the Lagrangian (3.18) in signs
of the fermionic mass terms.  The sign symmetry of the spectrum
$\n \leftrightarrow - \n$ is still ensured by the condition (3.10),
but now the zero mode emerges for even values of $s$, hence the
allowed for the validity of the linear solution (3.17) values of $s$
are the odd ones, for which $ \exp(4i \w x_1 ) = - \exp(4i \n x_1 )
$.  As a result, both in  even and odd cases the final spectral
equation for fermions is the same one, namely  $$ \exp(4i \n x_1 )= {
1- e^{-2ikd} {M-i (\n+k) \over M- i(\n-k) } \over  1- e^{-2ikd} {M+i
(\n-k) \over M+ i(\n+k) } }  \ { 1- e^{2ikd} {M-i (\n-k) \over M-
i(\n+k) } \over 1- e^{2ikd} {M+i (\n+k) \over M+ i(\n-k) } } \ .
\eqno(3.22)$$ It might seem, that the odd case is physically
unacceptable, since the Lagrangian (3.18) is not invariant with
respect to spatial reflection. However, it is caused by that
circumstance, that the considered models contain the minimal  set of
terms just to describe a single isolated bag.  It's easy to see that
the model, containing two odd bags, which are the mirror copies of
each other, should be P-invariant.  Therefore, both even and odd
bags possess actually the same physical status, and together cover
 all possible values of $s$ in Eq.(3.10) with the same spectral
eq.(3.22) for the eigenfrequencies $\n_n$.  In other words,
there exists a specific discrete symmetry in such systems, since
the most essential properties of the bag, e.g. the confinement of
fermions, properties of the intermediate region, the spectrum of
$\n_n$'s, etc., are invariant with respect to this change of signs in
the Lagrangian, which predicts merely the allowed values of $s$ in
Eq.(3.10).

Note also, that the Fig.1 does not present all the possible types of
the bags which could be obtained  combining the signs of the
chiral masses in all possible ways. In particular, there exist such
combinations for which the fermions should possess an energy
spectrum, which is different from that of Eq.(3.9),  and the
corresponding bags --- reveal a set of different physical
properties.  The study of this question   will be presented
separately [25].

\section{The total energy of the bag for the non-zero topological charge}

Without loss of generality we'll consider the bag
configurations with even values $s=2r$, which take place in
the P-invariant model (2.2). The resulting configuration of the
boson field reveals the structure shown on Fig.2.

\vskip 3 true mm
\epsfbox{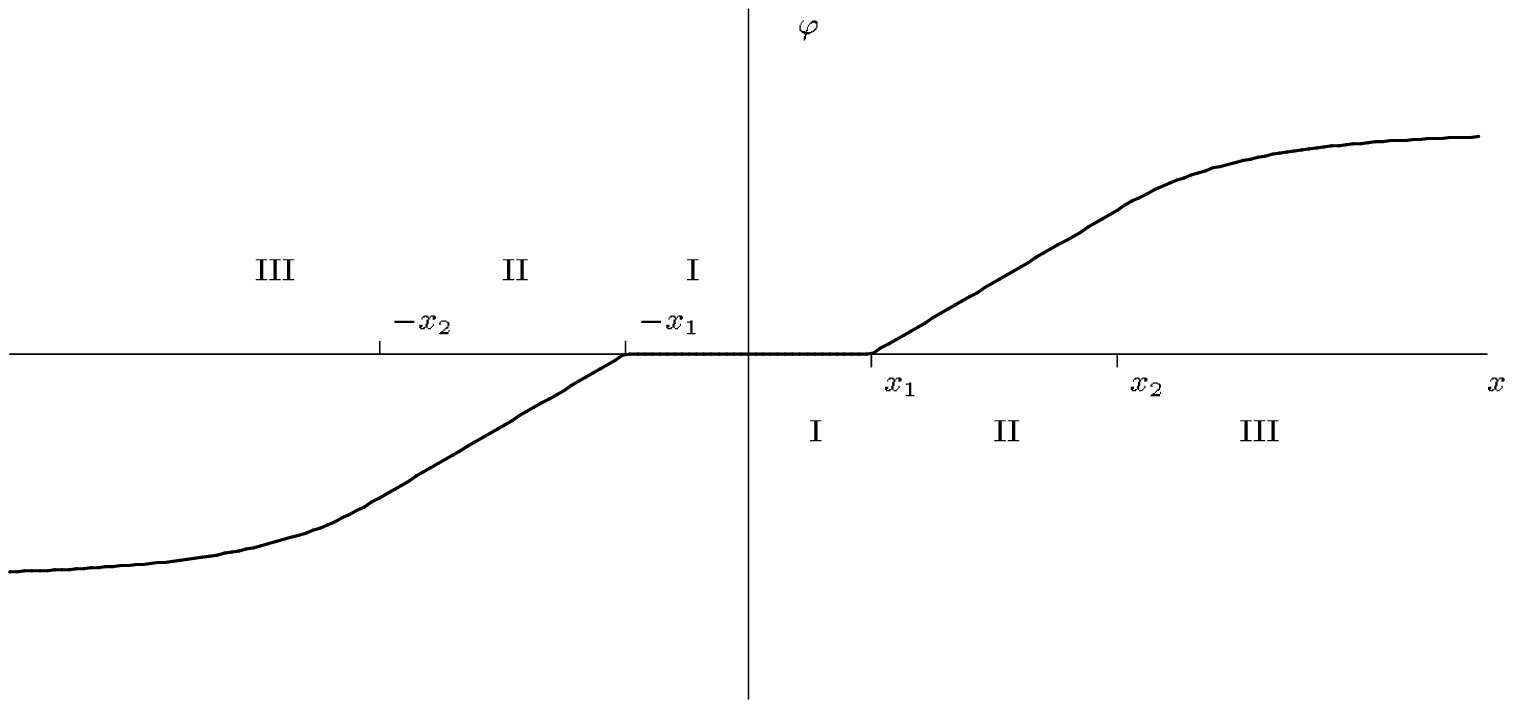}

\noindent Fig.2. The configuration of the boson field for an isolated
bag with non-zero topological charge.

\vskip 5 true mm

\noindent
In the region $I$ $\vf\equiv 0$, in
the region $II$ the boson field is the linear function (3.17), which
after restoring the $g$-dependence is sewn together
with the solution of Eq.(2.5b), which is responsible for the dynamics
of bosons in the bag's exterior.  To avoid the dependence on the
structure of self-interaction $V(\vf)$, we'll suppose, that in the
region $III$ the asymptotic expansion of the soliton solution of
Eq.(2.5b) for large $|x|$ can be used, namely $$ \vf_{sol}(x)={\pi
\over g} \(1- Ae^{-mx} \) \ , \quad x>x_2 \ , \eqno(4.1)$$ with $m$
being the (nonzero) meson mass in the bag's exterior, while for
$x<-x_2$ $ \vf_{sol}(x) $ is determined by oddness.  The chiral
symmetry in the exterior region of the bag is obviously lost.  It
should be noticed, however, that it occures due to specific features
of (1+1) D field theories, which make the presence of the meson mass
the necessary condition for the required soliton profile to be
formed.

The factor $\pi/g$ means that we deal actually with the
phase soliton with the total amplitude being multiple of $2\pi/g$,
since it is a period of the initial chiral interaction $\bp \exp
(i\g_5 g\vf)\p$.  The constant $A$ is determined from the continuity
conditions for boson field at points $x=\pm x_2$, what gives $$ x_1
={ r \over r+1 } (x_2 + 1/m) \ , \quad d={ x_2-r/m \over r+1 } \ .
\eqno(4.2)$$ The condition $d \geq 0$ yields then an additional
restriction for the size of the confinement region , i.e.  for the
external size $x_2$ of the bag $$ mx_2 \geq r \ ,$$ which shows, that
$r$ could be naturally interpreted as the index enumerating the
excited states of the bag, whose sizes increase with $r$.

Proceeding further, we find the relation between the
parameter $\l$ and the bag's size:  $$ 2\l = \pi m { r+1 \over
mx_2+1} \ , \eqno(4.3)$$ whence the total energy of the bosonic
soliton can be represented as $$E_{\vf} = m {\pi^2 \over g^2} {r+1
\over mx_2+1}.  \eqno(4.4)$$ The total energy of the bag is the sum
of $ E_{\vf}$ and of the fermionic contribution $E_{\p}$ $$ E_{bag}=
E_{\vf} + E_{\p} .  \eqno(4.5)$$ As it follows from
(4.4), the boson field energy  decreases smoothly for increasing
$x_2$  without producing any vacuum pressure, despite of the fact that
in the region $II$ the gradient of $\vf$ gives rise to the constant
positive contribution to the energy density $\1/2 \vf'^2=2\l^2$, what
could be identified with the vacuum pressure $B$ in the standard HCM.
Actually, it is an artifact of one spatial dimension in our problem:
when the bag's size increases, the gradient of $\vf$ in the region
$II$ decreases equally in any number of spatial
dimensions, while the volume of the region $II$ in one space
dimension increases only linearly and so cannot compensate the
decrease of $\l$, what would take place in 2- and 3- (space) D.
Thus, in (1+1)D all the non-trivial dependence of the total bag
energy $E_{bag}$ on the model parameters could originate from the
fermionic contribution $E_{\p}$ only, which is  the sum of the filled
Dirac's sea of negative energy states and positive energy valence
fermions $$ E_{\p}= E_{val} +E_{sea} \ .  \eqno(4.6)$$ For the ground
state of the bag described above, the sum (4.6) can be reduced to a
single universal expression by taking into account, that the charge
conjugation symmetry dictates the following definition of the Dirac's
sea energy [23, 26] $$ E_{sea}= \1/2 \sum \limits_{\w_n < 0} \w_n -
\1/2 \sum \limits_{\w_n > 0} \w_n \ .  \eqno(4.7)$$ If the
transformation from $\w_n$ to  $\n_n$ is sign preserving for all $n$
and so there are no valence fermions in the ground state of the bag
(to provide the vanishing v.e.v for the charge and chiral current),
one finds from (4.7): $$ E_{\p}=E_{sea}= \1/2 \sum \limits_{\n_n > 0}
 (- \n_n + \l) - \1/2 \sum \limits_{\n_n > 0} ( \n_n + \l) = - \sum
\limits_{\n_n > 0} \n_n \ . \eqno(4.8)$$ Note, that in (4.8) the
inequality is strict because there are no levels with $\n=0$.

If the parameter $\l$ appears to be
large enough, so that the initially negative level $\w_n=-|\n_n|+ \l
$ changes its sign, it turns into the filled valence state. The
 latter is again necessary  to provide the vanishing v.e.v for the
 charge and axial current. In this case it is convenient to calculate
$ E_{\p}$ in two steps. First, we consider the contribution from all
  states with $|\n_n|>\l$ to $E_{sea}$, which in analogy to (4.8)
  reads $$ E'_{sea}= - \sum \limits_{\n_m > \l } \n_m \ .
  \eqno(4.9)$$ To this expression the energy of emerging valence
fermions $E_{val}=-|\n_n|+\l $ and the contribution of the positive
levels with $\w_n=\pm |\n_n|+ \l $ to the Dirac's sea energy should
be added, what yields $$ E_{\p}=-|\n_n|+ \l - \1/2 [(-|\n_n|+
\l)+(|\n_n|+ \l)] + E'_{sea} = - \sum \limits_{\n_n > 0} \n_n \ ,
\eqno(4.10)$$  i.e. the same expression (4.8) as we have got for the
energy of fermions without filled valence states.

For what follows it is convenient
to introduce a set of new parameters, in terms of which the total
energy of the bag will be expressed in the most appropriate form.
First, we introduce the dimensionless quantities $$ \a=2Mx_1 \
  , \quad \b=2Md, \quad \r=2Mx_2 \ , \eqno(4.11)$$ and consider  in
more detail the Eq.(3.9), which determines the energy levels $\n_n$.
This equation has two branches of roots.  The first one corresponds
to real $k$ and in terms of parameters $\a$ and $\b$ can be
transformed into the following form $$ \tan \(\a \sqrt{1+x^2}\) = { x
\over \sqrt{x^2+1} } \ { x \cos \b x + \sin \b x \over 1- \cos \b x +
x \sin \b x } \ , \eqno(4.12)$$ where the unknown quantity is the
dimensionless $x$ defined through $k=Mx$, so that $\n=M
\sqrt{1+x^2}$. These real roots $x_n$ belong to the half axis $0 \leq
x_n < \infty $, since the fermionic wavefunctions are in fact the
  standing waves in a finite spatial box with degeneracy in the sign
 of $k$, while the corresponding frequencies $\n_n$ lie in the
interval $M \leq \n_n < \infty$.  The second branch corresponds to
imaginary $k=iMx, \ \n=M \sqrt{1-x^2} \ , \ 0 \leq x \leq 1$ and can
be derived from (4.12) by means of the analytical continuation $$
\tan \(\a \sqrt{1-x^2}\) = { x \over \sqrt{1-x^2} } \ { x \ch \b x +
\sh \b x \over \ch \b x + x \sh \b x -1 } \  .  \eqno(4.13)$$ For
this branch $0 < \n_n \leq M$.

Therefore, $\n_n$
 and so  $E_{\p}$ appear to be functions of two dimensionless
 parameters $\a$ and $\b$, which are not independent, however, but
 give in the sum the dimensionless total size of the confinement
domain $\r$:  $$ \a + \b = \r \ .\eqno(4.14)$$ Proceeding further, it
 is convenient to extract the mass of the "constituent quark" $M$
   from the sea energy and fermionic frequencies as a dimensional
factor:  $$ \ve_n=\n_n/M=\sqrt{1+x_n^2} \ ,\eqno(4.15)$$ hence $
E_{\p}= -M \sum \nolimits_n \ve_n$. Upon introducing the
dimensionless ratio of the two mass parameters of the model $$ \m = m
/2M \ , \eqno(4.16)$$ the dimensionless energy of fermions
$\E_{\p}=E_{\p}/M $ and analogously the dimensionless total energy
$\E_{bag}=E_{bag}/M $, for the latter one finds:$$ \E_{bag} = \E_{\p}
(\a, \b) + 2\m {\pi^2 \over g^2}   { r+1  \over \m \r+1 } \ ,
\eqno(4.17)$$ where the dimensionless parameters $\a, \ \b$ are
defined directly from $\m$ and $\r$ $$ \a ={ r \over r+1 }  \ (\r +
1/\m) \ , \quad \b={ \r-r/\m \over r+1 } \ .  \eqno(4.18)$$ So the
total energy of the bag depends ultimately on two dimensionless
parameters ---  $\m$ and $\r$, where the parameter $\m$ is fixed by
the ratio of the masses $m$ and $M$, while the optimal value of the
bag's size should be found from the condition of minimum of the total
 energy $\E_{bag}(\r)$ for given $\m$.

 To study the behaviour of
 $\E_{bag}(\r)$, first of all we have  to renormalize the fermion sea
energy $\E_{\p}$, which obviously diverges in the upper limit.
Let us start with the asymptotics of roots of Eq.(4.12) in the UV
 domain, when $x_n \gg 1$.  Representing the Eq.(4.12) as $$ \sin \a
\sqrt{1+x^2} = \1/2 (\sqrt{1+x^2} +x ) \sin \( \a \sqrt{1+x^2} + \b x
+ \d \) + \1/2 (\sqrt{1+x^2} -x ) \sin \( \a \sqrt{1+x^2} - \b x - \d
\) \ , \eqno(4.19)$$ where $\d = \arctan x \ , $ one finds that $$
\ve_n(\a, \b) = { \pi/2 + \pi n \over \r} + { (-1)^{n+1} \sin \[
(\pi/2 + \pi n ) \a/\r \] +1 + \b/2 \over \pi /2 + \pi n } + O
(1/n^2) \ .  \eqno(4.20)$$ In the expression (4.20) the first term
yields the quadratic and linear divergences in $\sum \nolimits_n
\ve_n $ and the second one produces the logarithmic one, while the
term with the sine doesn't yield  any divergence at all. To
 compensate the contribution of the first term, the energy of the sea
 of free fermions contained in the same "volume " $\r$ should be
subtracted, while the logarithmic divergence proportional to $\b/2$,
is compensated by the relevant one-loop  counterterm of the
boson self-energy [10]. The remaining logarithmic divergence
corresponding to the term $1/(\pi /2 + \pi n)$ doesn't depend on the
bag parameters, and originates from the fermion confinement inside
the bag, rather than by some local interaction.  Actually, it is the
(infinite)  energy of interaction between  fermions and the confining
potential (bag boundaries).  The appearance of such diverging surface
energy in $\E_{\p}$  is a specific feature of fermion vacuum
polarization in all the bag models [3,8,27-32].

In the considered 3-phase bag model this effect acquires some
additional features. First, it takes place for nonzero size $d\not=0$
of the intermediate phase  only, while the emerging surface
energy is negative and diverges as $\(-\sum \nolimits_n
1/(\pi /2 + \pi n)\)$. More concretely, if $\a \to \r$
then $(-1)^{n+1} \sin \[ (\pi/2 + \pi n ) \a/\r \] \to -1$, hence
there remains only the logarythmic term $\b/(\pi + 2\pi n)$ in the
asymptotics (4.20). Therefore in this limit $\E_{\p}$ becomes finite
just after subtraction of the energy of perturbative vacuum  and
adding the one-loop counterterm. On the other hand, the limit $\a
\to \r$ is equivalent to $\b/\a \to 0$, and so the infinite
interaction energy between fermions and bag boundaries takes place
for $d\not=0$ and the finite size of the central region (of the phase
of asymptotic freedom) of the bag only.

So  the considered 3-phase bag model doesn't actually reveal the
ability to the smooth transition into a 2-phase configuration for
$d \to 0$, although such an opportunity  exists formally on the level
of the initial Lagrangian (2.2). In fact, in the case of the
two-phase bag ($d \equiv 0$) the exact values of $\ve_n$ are
$\ve_n=(\pi/2 + \pi n)/\r$, and so $\E_{\p}$ becomes finite after
a single subtraction of the energy of the perturbative vacuum.
Therefore the transition between 2- and 3-phase bag configurations
requires an infinite amount of energy, what is a specific
feature of such many-phase systems. Note also, that the ultimate role
of the intermediate phase is the dynamical generation of the fermion
mass, whereas in the case of the two-phase bag ($d \equiv 0$) the
massless fermions are reflected directly from the bag boundaries.
From this point of view, there is an intimate connection between the
infinite surface energy of the bag and the circumstance, that for  $d\not=0$
the boundaries of the bag reflect massive fermions.

Within the considered 3-phase bag models we have an opportunity to
demonstrate this effect in a even more apparent way. For these
purposes let us consider a P-invariant model describing
the 1+1-dimensional analog of a "dibaryon", i.e. the
configuration with the  topological charge 2. Such
an object consists of two identical topological bags of the type
described above, which are placed so close to each other that their
neighbouring intermediate regions overlap. Upon dropping the
   $g$-dependence for simplicity, the corresponding Lagrangian reads
   $$\lagr = \bp i\hat\pd \p + \half (\pd_{\mu}\vf)^2 - {M \over 2}
   \[ \bp, e^{ig \g_5 \vf} \p \]_{-} \( \tt_{I} + \tt_{III} \)  - $$
   $$ - \half m_{0}^2 \( (\vf-\pi)^2 \tt_{II}^{(+)} + (\vf+\pi)^2
   \tt_{II}^{(-)} \) - \( {M_0 \over 2} \[\bp, e^{ig \g_5
   \vf}\p \]_{-}+V(\vf) \) \tt_{IV} \ , \eqno(4.21)$$ where
   $\tt_{I}=\tt (|x| \leq x_0) \ , \tt_{II}^{(\pm)}=\tt (x_0 <  \pm x
< x_1) \ , \tt_{III}=\tt (x_1 \leq |x| \leq x_2) \ ,
    \tt_{IV}=\tt(|x| > x_2)$.  The behaviour of the chiral fermionic
masses for this case is shown on the Fig.3.

\vskip 3 true mm
\epsfbox{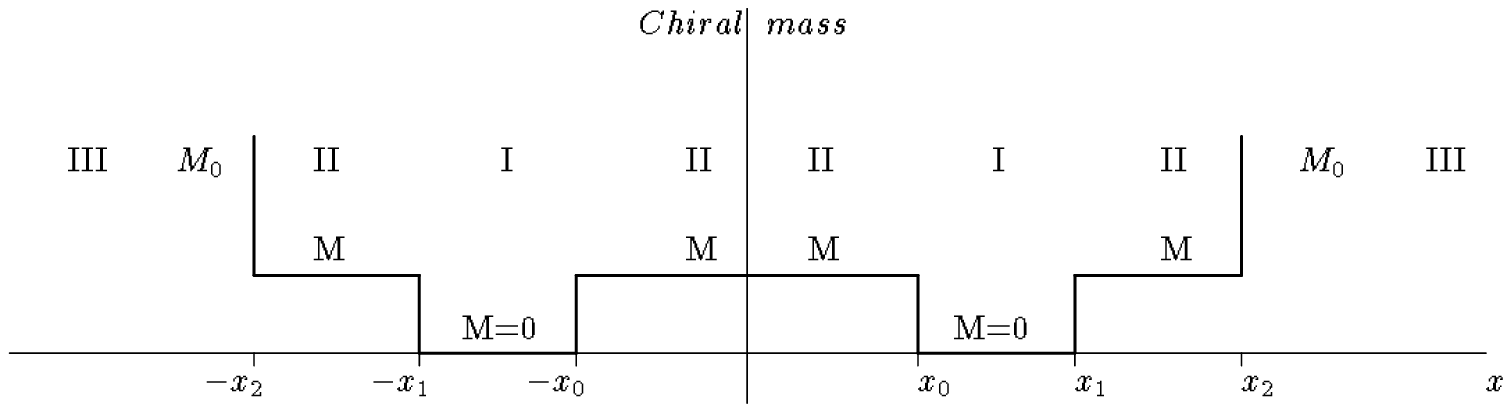}

\vskip -0.4cm
\noindent Fig. 3. The behaviour of the chiral fermionic masses for the
dibaryon configuration.

\vskip 5 true mm

\noindent
Again, we seek for the self-consistent solution of the model (4.21)
corresponding to such ''dibaryon'' configuration,  assuming
the linear behavior (3.4) for the boson field in the intermediate
regions and taking account of the sign-symmetry $ \n \leftrightarrow
-\n $ as  well as of the conservation of the chiral current $j_5=
\tilde j_5 $ for the transformations $ \c \to \tilde \c = \s_2 \c $.
Omitting some straightforward, but lengthy calculations, let us
present the main results.

The profile of the boson field corresponding to
the dibaryon configuration is shown on the Fig.4.

\vskip 5 true mm
\epsfbox{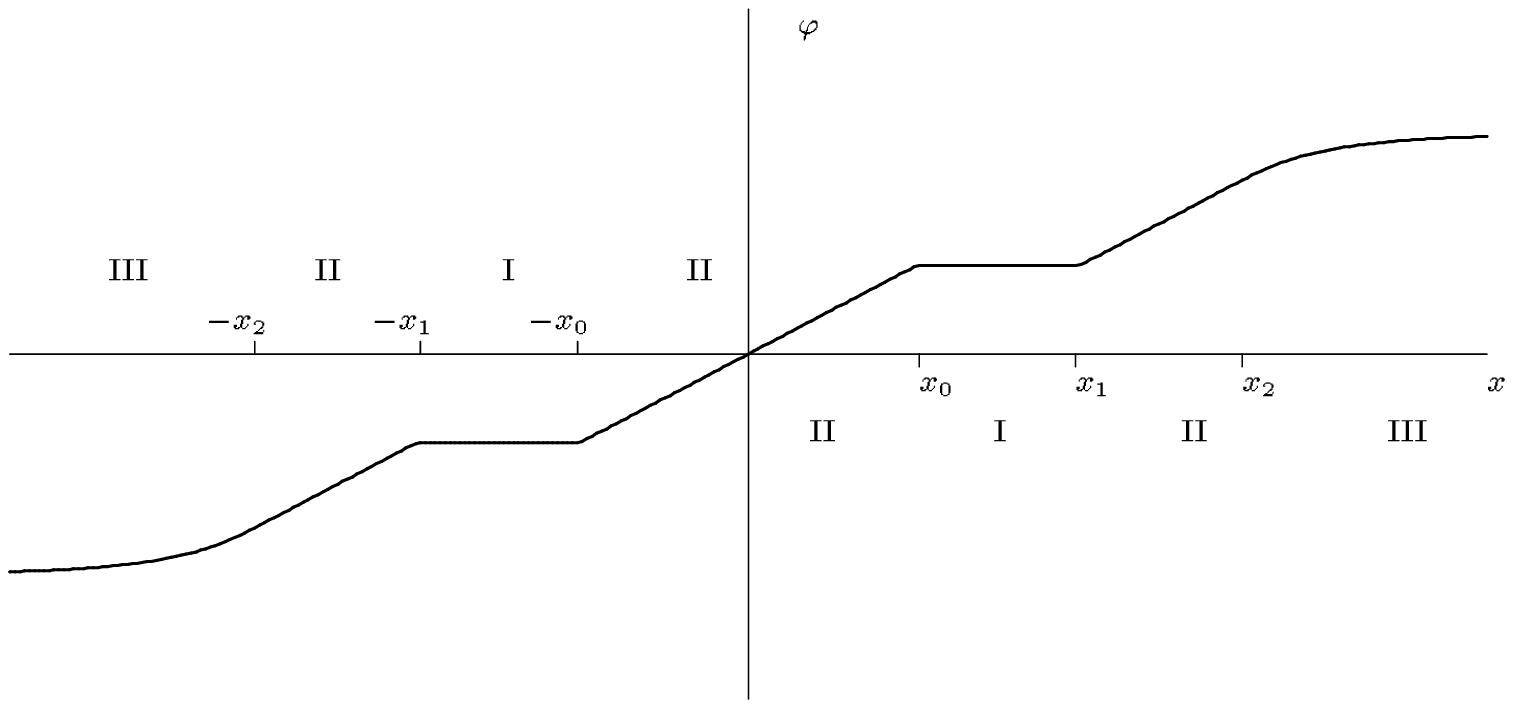}

\vskip -0.5cm
\noindent Fig.4. The boson field profile for the "dibaryon". 

\vskip 5 true mm
\noindent For the intermediate regions  of this dibaryon one obtains
$ \vf'=const=2\l \ , $ where $\l$ satisfies the condition $$ 2\l
a=\pi s \ , \quad  a=x_1-x_0 \ . \eqno(4.22)$$ The latter condition
is quite analogous to Eq.(3.10) for  a single isolated bag, because
the parameter $2x_1$ in (3.10), as well as $a$ in (4.22), has the
sense of the size of the region of asymptotic freedom of a single bag.
Note also,  that on the contrary to the single bag models considered
above, in the case of the dibaryon there are no zero modes in the
fermionic spectrum for any values of $s$. So there are no additional
restrictions imposed on the integer $s$ in Eq.(4.22).

It should be clear, that the 1+1-dimensional
model (4.21) cannot be considered as a  realistic model
of the dibaryon to any extent.  However, being simple and non-trivial
simultaneously, it turns out to be a very fruitful illustration for
the study of the origin of additional logarithmically divergent terms
$1/(\pi /2 + \pi n) $ in the UV-asymptotics of the fermionic
spectrum in such 3-phase bag models. The latter
 is again obtained from the corresponding
transcendent equation for  fermion levels, written in the
trigonometric form analogous to that of Eq.(4.12): $$ \sin \(2\a
\sq \) \( x \sq \cos \( (\b + \g) x + \d \) - x \cos \g x \) \ + $$
$$ +  \ \cos \( 2 \a \sq \) \( x^2 \sin \( (\b + \g) x + \d \) \ - \
\sq \sin \g x + \sin \g x \sin (\b x + \g) \) \ +  $$ $$ + \ \( \sq -
\cos (\b x + \g)  \) \sin \g x = 0 \ , \eqno(4.23)$$ where $ \a=Ma \
, \ \b=2Md \ , \ d=x_2-x_1 \ , \ \g=2M x_0 \ , \ \d= \arctan x $.
The parameter $d=x_2 -x_1$  is the size of the external
intermediate regions for each of the single bags forming the
dibaryon, while $2x_0$ is the size of their common internal
intermediate region, i.e.  the region of their mutual interaction.
In this case the UV-asymptotics of $\ve_n$'s has the following form
$$ \ve_n (\a, \b, \g) = { \pi/2 + \pi n \over \r} \ + $$ $$ + \ {
(-1)^{n+1} \( \sin \[ (\pi/2 + \pi n ) (2\a+\g)/\r \] - \sin \[
(\pi/2 + \pi n ) \g/\r \] \) +1 + (\b+\g)/2 \over \pi /2 + \pi n } +
O (1/n^2) \ , \eqno(4.24)$$ where $$\r=2\a + \b + \g = M( 2a + 2d +
2x_0) = 2M x_2 \ \eqno(4.25)$$ is the total dimensionless bag's size.
As in the case of a single isolated bag, the main divergent term in
the asymptotics (4.24) corresponds to the sea energy of free fermions
in the "volume" $\r$, while the logarithmic term, proportional to
$(\b+\g)/2$, is  exactly compensated by one-loop self-energy
counterterm.  The change of the coefficient in this term compared to
 (4.20) is caused by the fact, that in the considered case the
interaction between fermions and boson field takes place in the
region of the size $2d +2x_0$.  Besides this, there remains again a
logarithmically divergent term $1/(\pi /2 + \pi n)$, that corresponds
to the (infinite) energy of the interaction between fermions and the
confining  potential (bag boundaries),
and it follows from Eq.(4.24), that on the level of
divergent terms the surface energy of the dibaryon coincides exactly
with that of a single isolated bag.  So we
are led unambigously to the conclusion, that it is indeed the effect
of confinement of fermions in a simply connected region,  that yields
the term $1/(\pi /2 + \pi n)$ in Eqs.(4.20) and (4.24), since in
the dibaryon configuration the number of boundary points is just the
same as in the case of one isolated bag. Note also, that the direct
consequence of this statement is that in (1+1) D the dibaryon
configuration can't be obtained as a result of continuous  fusion of
two isolated bags, since when they are separated enough from each
other, the sum of their surface  energies is twice
larger than that of the dibaryon. In other words, in (1+1)D the
reconstruction of the bag's surface in the fusion-fission processes
requires an infinite amount of energy.

After all, it follows from (4.24), that for $\b \to 0$, i.e. for
vanishing external intermediate regions of the dibaryon one gets
$(-1)^{n+1} \sin \[ (\pi/2 + \pi n ) (2\a+\g)/\r \] \to -1$, what
 compensates the term $1/(\pi /2 + \pi n)$ , and  the infinite
interaction energy between fermions and bag boundaries disappears
too. So we are again led to the conclusion, made for a single bag by
analysis of the asymptotics (4.20), that the infinite surface energy
appears only then fermions pass through the intermediate phase just
before reflection from the bag boundaries.

As a result, for a 3-phase bag with $d\not =0$ the extraction of
the finite part from $\E_\p$ consists  actually  of two separate
procedures. The first one is the standard renormalization onto
perturbative vacuum with account of the one-loop counterterm,
caused by virtual fermion pairs [10]. The second one is the
compensation of the surface energy by means of an appropriate
subtraction, and both procedures suffer from an ambiguity in
the choice of subtraction point. In the "classical"  renormalization
scheme, the uncertainty in the choice of subtraction point is
cancelled by fixing the physical values for a corresponding number of
parameters. For obvious reasons, we avoid doing that in our
"toy" (1+1)D model, but instead  consider the most
straightforward approach to the compensation of divergences in the
sum (4.8), which preserves the continuous dependence of the
result of substraction on the model parameters. The essence of
this approach is that we subtract from $\sum \nolimits_n
\ve_n$ another sum with the same summation index $n$, whose common
term coincides exactly with the divergent part of
asymptotics (4.20). The result is the finite quantity $$ \tilde
{\E}_{\p}= -\sum \limits_n \[ \ve_n - \( {\pi/2+\pi n \over \r}+ {1
+\b/2 \over \pi/2 + \pi n} \) \] \ .  \eqno(4.26)$$ This method
requires no counterterms because all the divergences are already
cancelled by the subtracted sum. Of course, to some extent the
physical meaning of such procedure is lost. It should be
emphasized however, that it is only the (1+1)D case when the theory
with coupling $\lagr_I=G \bp (\s +i\g_5 \pi )\p$ is
(super)renormalizable and any counterterm has explicit physical
meaning. For higher space dimensions this is already not true and so
the procedure of compensation of divergences in the energy based on
(4.26) should not be considered as having no motivation. For more
detailed discussion on the extraction of the finite part from the
divergent Dirac's sea energy in 3+1 D HCM see Refs. [30-32].

Proceeding further, let us turn to the study of the total bag energy
$$ \E_{bag}=\tilde {\E}_{\p}(\a,\b) + 2\m {\pi^2 \over g^2} \ { r+1
\over \m\r+1 } \eqno(4.27)$$ as a function of the parameters $\m, \
\r$.  The analysis of the contribution of convergent logarithmic part
from the sine-term in the asymptotic expression (4.20) to $\tilde
{\E}_{\p}$ yields the first feature of $\E_{bag}$. Let us transform
this contribution to the form $$ \(\tilde {\E}_{\p}\)_{log} (\a,\b)=
 { 1 \over \pi} \sum \limits_{n \gg 1} (-1)^{n} { \sin \[ (\pi \a /
\r) ( n + 1/2 ) \]  \over  n + 1/2  } \ \eqno(4.28)$$ and then use
the well-known relation $$ \sum \limits_{n=0}^{\infty}  (-1)^{n} {
\sin \[ z ( n + 1/2 ) \] \over  n + 1/2  } = \ln \tan (\pi/4+z/4) \ ,
\quad |z|<\pi \ .  \eqno(4.29)$$ It is easy to see, that the sums
(4.28) and (4.29) possess the similar common term, while the sum
(4.28) diverges as $(- \ln (\pi - z))$ when $z \to \pi$.  Hence, for $
\pi \a /\r \to \pi   $ , what means $\b \to 0$,  the sum (4.29) will
show the similar behavior, namely: $$ \( \tilde {\E} _{\p}\)_{log}
(\a,\b) \to  - {1 \over \pi} \ln \b \ , \quad \b \to 0 .
\eqno(4.30)$$ Therefore, both the regularized fermion energy (4.26)
and the total bag's energy reveal the logarithmic singularity for
$\b \to 0 $, i.e. for $\r \to r/\m$, what is in
complete agreement with the qualitative analysis of the vacuum
polarization effects in the 3-phase bag, performed above.  In
particular, for $\b \to 0$ both the intermediate phase and
the infinite energy of coupling between fermions and bag boundaries
disappear, what in the present case shows up in the unlimited
increase of fermion energy, since after the renormalization of
$\E_\p$ by means of subtraction (4.26) the divergent part of the
surface bag energy is included in the onset of energy.

$\E_{bag}$  will also grow for $\r \to \infty$.  In this case
$\a/\r \to r/(r+1)$, so the logarithmic term (4.28) remains finite,
  what implies, that we have to deal now with the whole sum (4.26).
However, the leading order behaviour of $\tE_\p$ could be
evaluated  from (4.26) quite effectively by using the fact that for
$\r \to \infty$ the fermionic spectrum becomes quasicontinous, what
allows to transform the sum over $x_n$ into integral over $dx$.  In
particular, the analysis of distribution of the roots of Eq.  (4.12)
shows, that for this limit $\sum_n \ve_n$ is approximated by the
 following (divergent) integral:  $$\sum_n \ve_n \to {1 \over \pi} \
\int \! dx \ \sqrt{1+x^2} \ \[ \b + {1 \over 1+x^2} + \a {x^2 \over
x^2+ \sin^2 \(\a\sqrt{1+x^2}\)} - { \sin\(\a\sqrt{1+x^2}\)
\cos\(\a\sqrt{1+x^2}\) \over  \sqrt{1+x^2} \( x^2+ \sin^2
\(\a\sqrt{1+x^2}\) \) } \] \  .  \eqno(4.31)$$ For the subtracted sum
in (4.26) one can easily find $$ \sum \limits_n \( {\pi/2+\pi n \over
\r}+ {1 +\b/2 \over \pi/2 + \pi n} \) \to  {\r \over \pi} \ \int \!
dx \  \( x+ {1+\b/2 \over \r x} \)  \ .  \eqno(4.32)$$ The integrals
(4.31) and (4.32) have obviously the same divergent part $$ {1 \over
\pi} \ \int \! dx (\r x + 1/x + \b /2 x) \ , $$ and so their
difference yields a converging integral, in agreement with the
subtraction procedure. The leading term of the integrand in this
difference, taken with the (correct) inverse sign, is $\b/8\pi x^3$.
Since $\b/\r \to 1/(r+1)$ for $\r \to \infty$, this finally leads to
the emergence of the positive, proportional to $\r$, contribution to
$\tilde {\E}_{\p}$, and correspondingly to $\E_{bag}$.

The numerical calculation confirms completely such qualitative
predictions for the behavior of $\E_{\p}(\r)$ and $\E_{bag}(\r)$. In
the present paper such calculation has been performed for
$\m=0.25$, what corresponds approximately to the ratio
$m_{\pi}/2M_Q$, where the constituent quark mass $M_Q$ is assumed to
be equal to 300 MeV, and for $g=1$, because the energy of boson
soliton doesn't have any significant influence on the main properties
of $\E_{bag}(\r)$. The results of $\E_{bag}(\r)$  calculation for
$r=1,2,3,4,5$ are depicted on the Fig. 5 and show, that
the size and energy of the solution, determined from  the minimum of
$\E_{bag}(\r)$, continuously grow  for increasing $r$,
whereas the curvature of $\E_{bag}(\r)$ in the minimum decreases,
what provides an unambiguous interpretation of configurations with
$r>1$ as excited states of the bag.

\section{Bags with zero topological charge}

Now let us consider the bag with vanishing topological charge,
what for the relevant configuration of the
boson field should be an even one $\vf(x)=\vf(-x)$. The principal
difference between this case and the previous one is that for even
$\vf(x)$  the sign symmetry $\w \leftrightarrow -\w$ is an immanent
feature of the spectral problem for fermions (2.7-10),  what can be
easily justified by means of the following transformation of
fermionic wavefunctions $$ \p_\w(x) \to \p_{-\w}(x)= \pm \g_5
\p_\w(-x) \ .  \eqno(5.1)$$ However, the corresponding axial currents
are related now in the following way $$ j^5_{-\w}(x)=-j^5_\w (-x) \ ,
\eqno(5.2)$$ so there is no automatic compensation between positive-
and negative-frequency terms in the v.e.v. of $J_5(x)$. From (5.2)
one can derive only the relation $$ \<J_5(x)\>_{sea} =
\<J_5(-x)\>_{sea} \ , \eqno(5.3)$$ what guarantees the consistence of
Eq.  (2.4b) with respect to parity. The direct consequence of such
fermion properties is that the even
 configuration of the boson  field, similar to (3.17),
 $$ \vf(x)= \Bigg\{ {+2 \l (x-x_1) \ , \quad \ \ x_1 \leq x \leq x_2
\ , \atop \ \ -2\l (x+x_1) \ , \quad -x_2 \leq x \leq -x_1 \ , }
\eqno(5.4)$$ is no longer an exact solution of Eqs.  (2.4), since in
this case $\<J_5(x)\>_{sea} \not \equiv 0$   in the region $II$.

Nevertheless, the configuration (5.4) plays an important role in the
study of  the non-topological case. First of all, for small values of
$g$ it turns out to be a rather good approximation to the
precise solution.  To argue this statement, let us note firstly, that
the substitution $\vf = \tvf/g$  removes $g$ from the Eq.(2.4a),
while Eq.(2.4b) will contain $g$ only as a coefficient in the
r.h.s, namely $$ \tvf ''= ig^2 {M \over 2} \< \[\bp, \g_5 e^
{i\g_5\tvf}\p\]_{-}\> \ .  \eqno(5.5)$$ Proceeding further,
it would be natural to assume that the potential
$V(\vf)$ depends on $g$ as $$ V(\vf)= W(g\vf)/g^2 \ , \eqno(5.6)   $$
where $W(f)$ should be an even polynom to preserve
the (anti)symmetry of soliton solutions.  Therefore, for
small $g$ there emerges  quite naturally  an expansion
in powers of $g^2$ in the problem.
 Within this expansion, the zero-order approximation for the
rescaled boson field $\tvf (x) $  is the
configuration (5.4) in the region $II$, while $\tvf(x) \equiv 0$ in
the region $I$,  and in the region $III$ it is assumed to merge
with the asymptotics of an even soliton solution $$
\tvf_{sol}(x)=\pi \(1- Ae^{-m|x|} \) \ , \quad |x|>x_2 \ ,
\eqno(5.7)$$ quite similar to the topological case.
Sewing together (5.4) and (5.7) by means of the continuity
conditions for $\vf$ and $\vf'$ yields the following relation
$$ 2\l =  {\pi m \over md+1} \ , \eqno(5.8)$$ whence
for the energy of the boson field one finds  $$ E_{\tvf}= {\pi^2 m
\over md+1} \ . \eqno(5.9)$$  Note, that in the initial variables
$\vf$ the dependence on $g$ in $ E_\vf $  is restored by adding the
coefficient $1/g^2$.

Now let us show,  that the first-order  $O(g^2)$ correction   to the
energy of the boson field (5.9) vanishes exactly for any current in
the r.h.s. of Eq.(5.5), provided the asymptotics (5.7) for the boson
field in the region $III$ remains valid beyond the perturbation
expansion in $g^2$, what implies, that  the corrections caused by the
r.h.s.  of (5.5) disturb only the value of the parameter $A$.
To simplify calculations, we will consider further only the positive
half-axis. The contribution of the negative one is
exactly the same.

From the asymptotics (5.7) we derive  $$ m\tvf
(x_2) + \tvf'(x_2)=\pi \ , \eqno(5.10)$$ while in the region $III$ $$
\tvf'_{III} (x)= \tvf'(x_2) e^{-m(x-x_2)} \ , \eqno(5.11)$$
which are valid beyond the $g^2$-expansion as well. Using the virial
theorem which is also relevant beyond the latter expansion,
 we obtain the following general expression for the
contribution of the region $III$  to $ E_{\tvf} $ $$
{E_{\tvf}}_{III}= \int \limits_{III} \! dx \ \tvf'^2 = {
\(\tvf'(x_2)\)^2 \over 2m } \ .  \eqno(5.12)$$ Proceeding further,
on account of the first-order correction from the non-zero $
\<J_5(x)\> $ one obtains  for the boson field in the region
$II$ $$ \tvf(x)= 2 \l (x-x_1) + g^2 \tvf_1 (x) \ . \eqno(5.13)$$
At the same time,  it follows from the condition $\tvf(x) \equiv 0$
in the region $I$ and the boundary conditions (5.10) that $$
\tvf_1(x_1)=0 \ , \quad m\tvf_1(x_2) + \tvf_1'(x_2)=0 \ .
\eqno(5.14)$$ Then for the  boson field energy in the region $II$
with the first  $O(g^2)$ correction one finds $$ {E_{\tvf}}_{II}=\1/2
\int \limits_{II} \! dx \ \tvf'^2 = 2 \l \(\l d + g^2 \tvf_1 (x_2) \)
\ .  \eqno(5.15)$$ On the other hand,  it
follows in the same  approximation from (5.12) and (5.13), that $$
             {E_{\tvf}}_{III}= { 2 \l \over m} \( \l + g^2 \tvf'_1
(x_2) \) .  \eqno(5.16)$$  Returning to Eq.(5.14), one finds,
that in the sum $ {E_{\tvf}}_{II} + {E_{\tvf}}_{III} $ the
contribution of $\tvf_1$ vanishes.  In other words, within
the $g^2$-expansion the corrections to the leading
approximation (5.9) in $ E_{\tvf} $, caused by
nonvanishing $ \<J_5(x)\> $, begin from the second order $O(g^4)$
only.

Next, let us note, that for fermions the
$g^2$-expansion starts from the order $O(g^0)$. In this
approximation the spectral problem (2.7-10) leads to
the following equation for the fermionic spectrum $$ \exp(4i\w
x_1)=\left[ \frac{\nu_{-}+k_{-}}{\nu_{+}+k_{+}} \right]
\frac{1-e^{-2ik_{+} d}\frac{M-i(\nu_{+}+k_{+})}{M-i(\nu_{+}-k_{+})}}
{1-e^{-2ik_{+} d}\frac{M+i(\nu_{+}-k_{+})}{M+i(\nu_{+}+k_{+})}}
\frac{1-e^{2ik_{-} d}\frac{M-i(\nu_{-}-k_{-})}{M-i(\nu_{-}+k_{-})}}
{1-e^{2ik_{-} d}\frac{M+i(\nu_{-}+k_{-})}{M+i(\nu_{-}-k_{-})}},
\eqno(5.17)$$ where $\nu_{\pm}=\w \pm \l, \quad \nu_{\pm}^2=
 k_{\pm}^2+M^2$.
The total energy of the bag is still given by the sum (4.5), where
the fermion energy has the following form
$$ E_\p=- \sum \limits_{\w_n < 0} \w_n \ . \eqno(5.18)$$
In (5.18), as well as in (4.8), the inequality $\w_n <0$ is strict,
 because for the configuration (5.4) there are no levels with
$\w_n=0$ for any values of $x_1 , x_2$.

Finally, after restoring
dependence on $g^2$ in $ E_\vf $ we obtain the following expression
for the total energy of the bag $$ E_{bag}={\pi^2 \over g^2} \ { m
\over md+1} + E_\p + O(g^2) \ , \eqno(5.19)$$ where the first two
leading terms in $E_{bag} $ ---  the bosonic $O(1/g^2)$ and fermionic
$O(g^0)$ --- are determined by the zero-order approximation for the
boson field (5.4.7) only, while the corrections start
with $O(g^2)$ terms,  simultaneously in
the bosonic and fermionic parts of the total energy.

The numerical calculation completely confirms such behaviour of the system
for small $g$. On the Fig.6 a typical profile of the numerical
solution for the  boson field is shown, obtained by means of
minimization procedure of the total energy functional for the values
of $g$ from the interval $0.01 \div 0.5$, which shows clearly
that for such values of $g$ the exact solution in the
intermediate region is  almost indistinguishable from the linear
function (5.4).

\vskip 3 true mm
\epsfbox{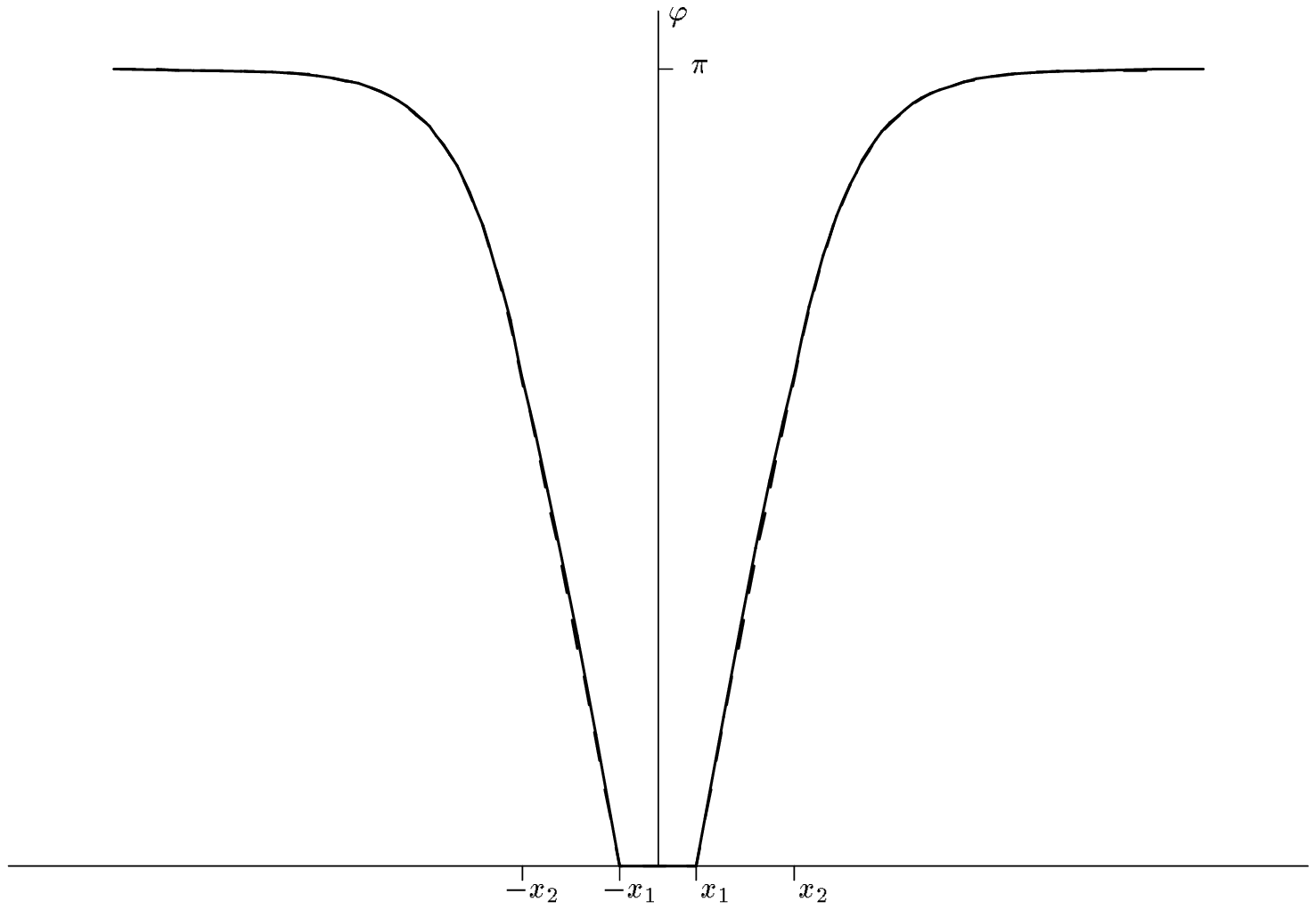}

\noindent
\vskip 0.5cm
Fig.6. The profile of the numerical solution for boson field,
obtained by the minimization procedure of the bag's total energy
functional for $g=0.1$ on a spatial lattice with $\simeq 10^3$
nodes. The self-interaction of bosons  in the region $III$
is taken in the form $V(\vf)=\(\pi^2 -(g\vf)^2\)^2/2g^2$.
The dashed line, which
 merges almost with the solid one,  corresponds to the calculation
with the doubled number of nodes on the lattice, what indicates
the accordance of the numerical solution with the continuous limit.

\vskip  1 true cm
\noindent
Moreover, the respective simplicity of Eq. (5.17) makes it
possible to analyse the fermionic spectrum in a semi-analytical way,
what in turn allows to use the configuration (5.4,7) as  a trial one
for a qualitative study of the ground state properties for the
non-topological bag  for even larger values of $g
\simeq 1 $.

 Thus, in further analysis of the main properties of the
 non-topological bag we will use the first two terms in the total
energy (5.19), which can be found directly from the
configuration (5.4,7). Recalculating $E_{\vf}$ to the
dimensionless variables, introduced in (4.11,14-16), one obtains $$
\E_{bag}(\a,\b) = \E_\p (\a,\b)+ {\pi^2 \over g^2} {2 \m \over \m\b+1} \ .
\eqno(5.20)$$ Note, that in (5.20) the sum of the parameters $\a, \b$
defines the dimensionless size of the bag $\r$, but on the
contrary to the previous case  in all other respects they should be
considered as independent ones. So the study of the bag's energy
as a  function of its geometry becomes a qualitively different
problem of finding the two-dimensional surface $ \E_{bag}(\a,\b) $.

The extraction of the finite part from $\E_\p(\a,\b)$  undergoes the
same main stages as in the topological case, but reveals some
peculiar features, caused by the independence of  $\a$ and
$\b$.  After some transformations it's not difficult to obtain from
(5.17)  the UV-asymptotics of the energy levels in the following form
 $$ \ve_n (\a,\b)= { \pi/2 + \pi n \over \r} + { (-1)^{n+1} \cos(2\l
d) \sin \[ (\pi/2 + \pi n ) \a/\r \] +1 + \b/2 \over \pi /2 + \pi n }
+ O (1/n^2) \ .  \eqno(5.21)$$
It follows immediately from the structure of the logarithmic term in
(5.21), that as in the topological case, the renormalized via
asymptotics $\tE_\p$, as well as $\E_{bag}$, show up a logarithmic
divergence $(-\ln \b /\pi)$ for $\b \to 0$.  Besides this, $\tE_\p$
and $\E_{bag}$ increase for $\b \to \infty$ and finite $\a$.  Since
$\r$ grows simultaneously with $\b$, this effect turns out to be
quite similar to the increase of $\tE_\p$ and $\E_{bag}$ for $\r \to
\infty$ in the topological case:  in the UV-domain the difference
between $\n_+$ and $\n_-$ vanishes and Eq.(5.17) turns into (3.9).
Thereafter to analyse the renormalized $\tE_\p$ one may use the
integral approximation (4.31,32), in which for $\b \to \infty$ the
main term in the integrand of $\tE_\p$ is positive and proportional
to $\b$.

The behaviour of $\E_\p$ and
$\E_{bag}$ for $\a \to \infty$ and finite $\b$ requires
special consideration, for in this case the logarithmic term
with the sine in asymptotics (5.21) becomes significant again,
but unlike the case of $\b \to 0$,  there appears now an additional
factor $\cos (2\l d)$. Since $2 \l d = \pi \m \b /(\m \b
+1)$, the sign of this multiplier can be either positive or negative
depending on the current value of $\b$.  However, in this limit there
becomes significant another effect, namely the proportional to
$\a$ increase of the number of levels on those branches of
the fermionic spectrum, which correspond to the imaginary values of
$k_{\pm}$ in (5.17),  where $0 < \n_{\pm} \leq M$. Directly for the
Eq.(5.17) this effect shows up in an intricate  enough way  due to
the presence of separate branches for imaginary $k_+$ and $k_-$ and
therefore can be analysed in detail only numerically, but its essence
could be understood quite simply, if we  neglect for a while the
difference between $k_+$ and $k_-$.  Then we are left with only one
branch with $0 < \n_n \leq M$ determined from Eq. (4.13), from which
it is easily seen that for $\a \to \infty$ the spectrum of energy
levels belonging to this branch becomes quasicontinous with the
interval between the levels of order $\pi/\a$, hence $\sum_n \ve_n$
over this branch can be approximated by the integral $$ \sum
\limits_{ n \atop II \ branch } \! \ve_n \to - {1 \over \pi} \
\int_0^1 \!  dx \ \sqrt{1-x^2} \ \[ -{\a x \over \sqrt{1-x^2} } + { x
\( \b +  1 / ( 1-x^2) \) -  \sh (\b x + \g) / \sqrt{1-x^2} \over  \ch
(\b x + \g ) - \sqrt{1-x^2}  } \] \ . \eqno(5.22)$$ Therefore, for
$\a \to \infty$ the second branch contribution to $\sum_n
\ve_n$  takes the form $\a/2\pi \ +$ finite terms depending
on $\b$ only.  Transforming further the subtracted sum to the
integral one obtains $$ \sum \limits_{ n \atop II \ branch } \(
{\pi/2 + \pi n \over \r} + {1+ \b/2 \over \pi/2 + \pi n } \) \to { 1
\over \pi} \ \int \limits_{\pi/2}^{\a} \!  dx  \ \( {x \over \r} +
{1+ \b/2 \over x} \) \ , \eqno(5.23)$$ whence it follows that for $\a
\to \infty$ the main terms in the subtracted sum should be  $${\a^2
\over 2 \pi \r} + { 1+ \b/2 \over \pi} \ln \a \ .  \eqno(5.24)$$
The leading, proportional to $\a$ terms in Eqs.(5.22,24) cancel each
other, so after subtraction the contribution of the second branch to
the renormalized $\tE_\p$ should be $ (1/\pi) (1+ \b/2) \ln \a \ +$
finite terms. For the case of separate branches for $k_+$ ¨ $k_-$
the general features of their asymptotic behaviour for $\a \to \inf$
remain the same.  As a result, after combining this asymptotics with
the corresponding input of the logarithmic term in the UV-asymptotics
(5.21), for the leading term in $\E_\p$ for $\a \to \infty$ one finds
$$ {1 \over \pi} \( 1+ \b/2 + \cos 2\l d \) \ \ln \a \ , \quad \a \to
\infty \ , \eqno(5.25)$$ which is definitely positive for all $\b$.
So in this limit the bag's energy also grows, but now proportional to
$\ln \a$.

The numerical calculation confirms
completely such qualitative behaviour of $\tE_\p$  and $\E_{bag}$. It has been performed for the same
values of $\m=0.25$ and $g=1$ as for the topological bags,
shown on Fig. 5. The most specific feature of the non-topological
case is that in accordance to (5.9) the boson field energy
depends now  on $\b$ only, while the renormalized energy of fermions
decreases smoothly for $\a \to 0$. A profile of the surface
$\tE_\p (\a, \b)$ is depicted on Fig. 7, where
$\tE_\p =0$ for $\a=\b=0$ is an artifact of the
chosen subtraction method. As a result, there isn't any non-trivial
minimum in the total energy for the non-topological case at all, while
 the minimal energy is achieved by the configuration with
vanishing size of the phase of asymptotic freedom and for
finite non-zero $\b$, what is clearly seen from Figs. 8,9, on which
the profiles of the 2D surfaces $\E_{bag}(\a, \b)$ are presented in
different scales. So for the bags with zero topological (i.e.
baryon) charge the considered three-phase model predicts that the
main role should be played by the intermediate phase of constituent
quarks, what is quite consistent with semi-phenomenological
quark models of mesons [9,33].

\section{Conclusion}

This work  was aimed at the construction of a
reasonable model of a hybrid chiral bag without special assumptions
about the intimate relations between the fermionic and bosonic
phases.  Our results
show that such a model can be actually formulated in a quite
consistent fashion, and to certain extent could be more efficient
way of description of low-energy hadron physics compared to the
standard HCM [1-8]. The main advantages of such an approach include,
first of all, a more correct formulation of chiral boundary
conditions for which all the components in the Lagrangian possess
clear physical meaning, the existence of the intermediate phase,
describing quasifree massive "constituent quarks", as well as a quite
acceptable physically behaviour of the total bag's energy as a
function of its size, which takes the form of an infinitely deep
potential well with a distinct minimum in the topological case,
whereas in non-topological case the minimal energy of the bag is
achieved by the configuration, where the phase of asymptotic
freedom disappears.

Besides this, in this model the condition of fermions confinement,
incorporated into it from the very beginning, shows up more
explicitly.  It manifests, in particular, in the fact that there is
no need in the vacuum-pressure term $B$, which in the standard
approach is inserted into the model by means of some extra
assumptions, since in the considered case the Dirac's sea
polarization itself produces the  infinite increase
of energy at large distances. Another essential feature is the
appearance of infinite interaction energy between fermions and bag
boundaries (confining potential) for $d \not=0$, what
implies, that the size of the intermediate region doesn't actually
vanish, although on the level of the initial Lagrangian the formal
limit $d \to 0 $ exists and yields the standard two-phase model of a
hybrid bag. In other words,  such a 3-phase model cannot be
continously transformed into a 2-phase one, what is the ultimate
reason of remarkably different features of this model compared to
the standard 2-phase ones.

It is worth-while to mention once more the  question of
the choice of method of calculation of the Dirac's
sea averages for fermion bags. The method we used is
based on the discreteness of the fermionic energy spectrum what by
means of quite obvious considerations leads to very simple
solution of self-consistent equations of the bag in the intermediate
region. Let us remark, however, that despite of arguments in
favour of such method of calculation of sea averages, we cannot
completely reject alternative methods like the
thermal regularization. The question of which one is more adequate to
the physics of the problem, should be answered only by means of
detailed study of realistic models.

It should be also emphasized, that by constructing such a 3-phase
model we have substantially  used the condition of
lorentz-covariance.  The initial formulation of the
model is a local field theory and regardless on the variety of
classical solutions one needs to deal with, the covariance is broken
only spontaneously,  and so can be freely restored by means of
methods of Refs.[21] based on the covariant group
center-of-mass variables for a localized quantum-field system.
 However, such an explicitly covariant framework  requires
some essential changes in the calculation techniques, since the
invariant dynamics of fields in the c.m.s. acquires a specific
finite-difference form [21], and so will be considered separately.

This work was supported in part by RFBR under Grant 00-15-96577
and by Sankt-Petersburg Concurrency Centre of Fundamental Sciences,
Grant 00-0-6.2-22.

\vskip 2 cm

\epsfbox{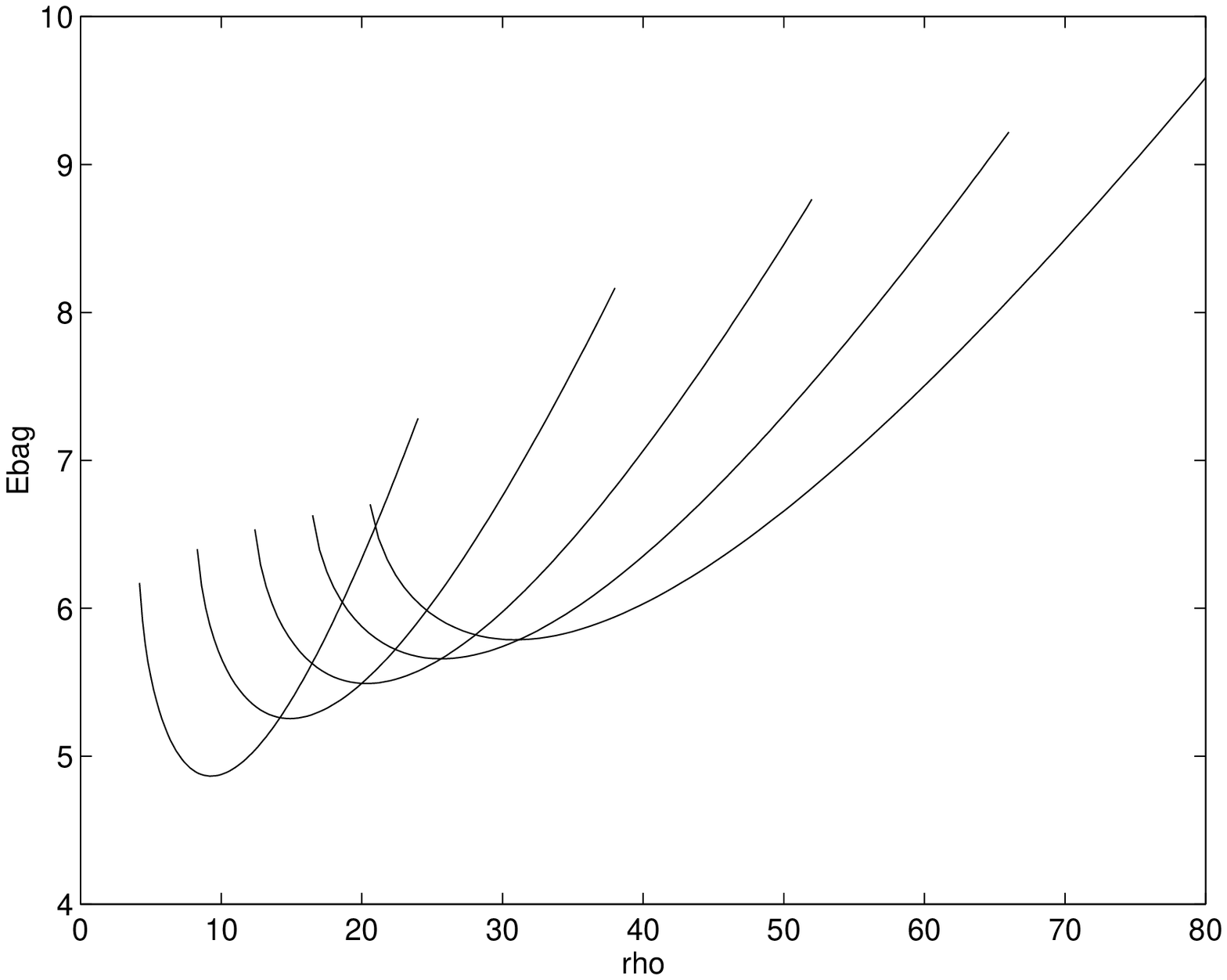}

\vskip 0.7cm
Fig.5. $\E_{bag}(\r)$ for $r=1,2,3,4,5$ in the topological case.

\eject
\epsfbox{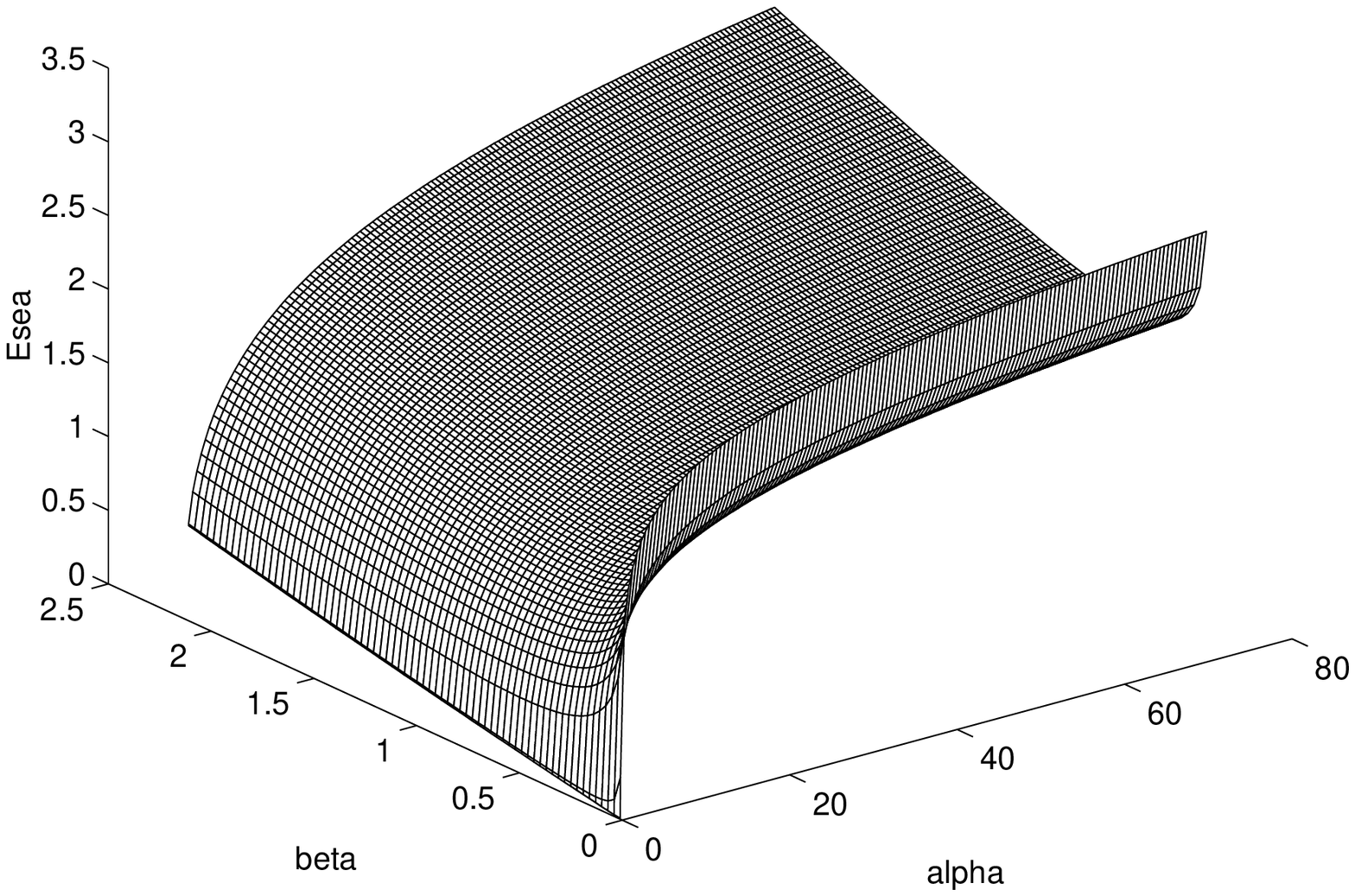}

\vskip 0.7cm
Fig.7. The profile of the surface $\tilde{\E}_{\p}(\a, \b)$ for the 
non-topological bag.

\eject
\epsfbox{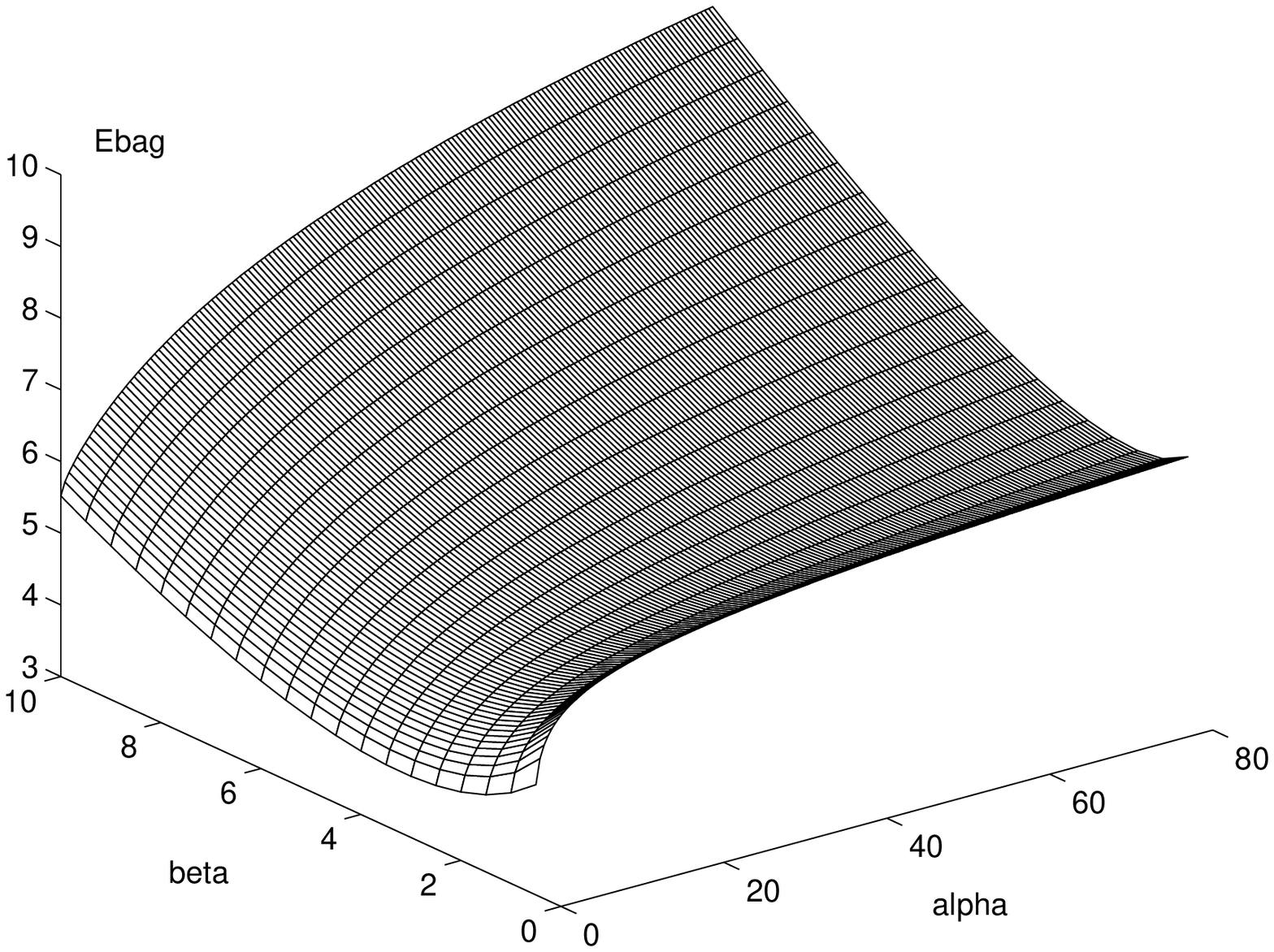}

\vskip 0.7cm
Fig.8. The profile of the surface $\E_{bag}(\a, \b)$ for the 
non-topological bag.

\eject
\epsfbox{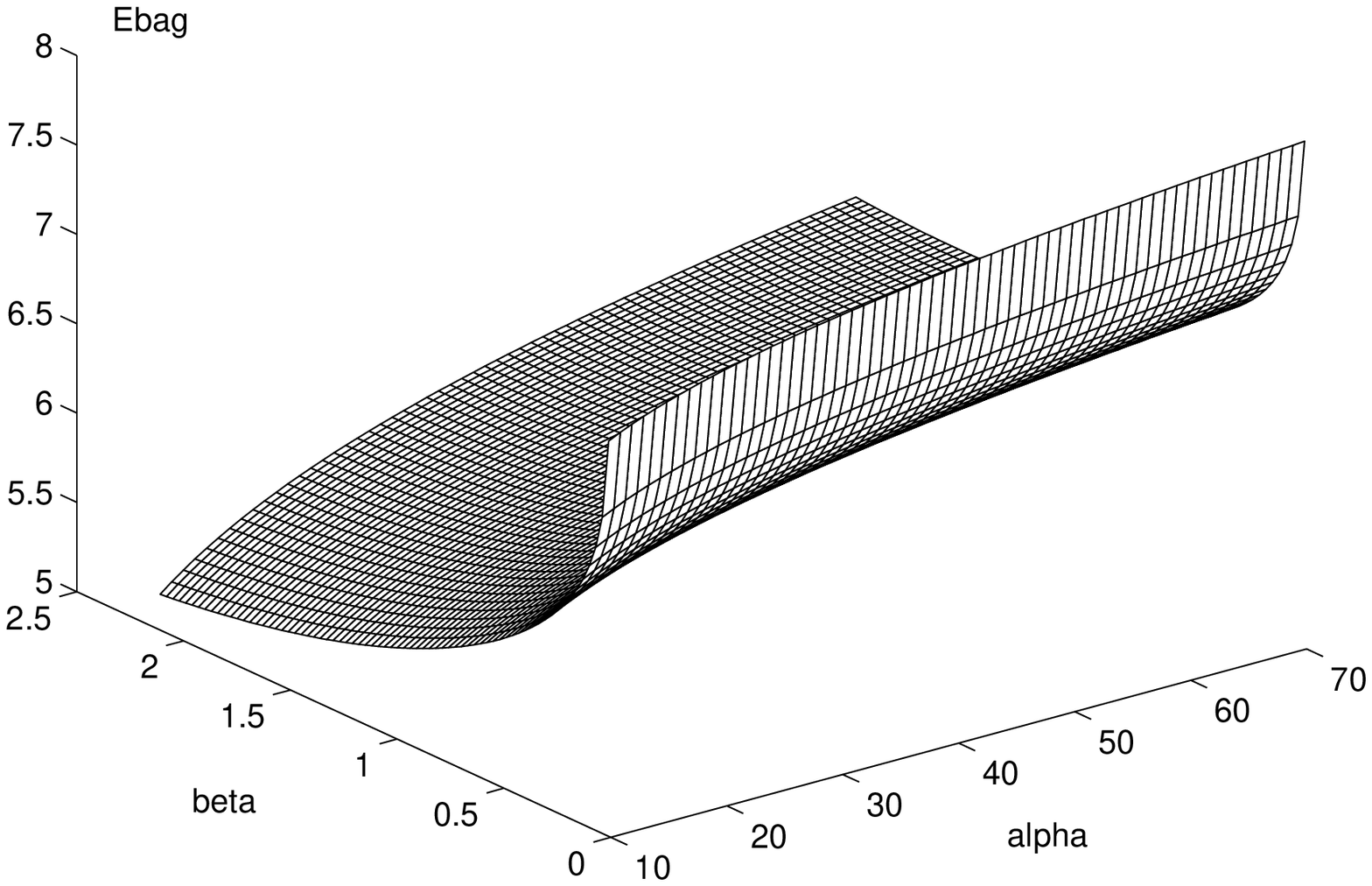}

\vskip 0.7cm
\noindent Fig.9. The profile of the surface $\E_{bag}(\a, \b)$ for the 
non-topological bag, rescaled to observe the behavior for small $\b$.
\eject

\end{document}